\documentclass[a4paper,fleqn,usenatbib]{mnras}
\usepackage{soul}
\usepackage{epstopdf}
\usepackage{longtable}
\usepackage{pdflscape}
\usepackage{graphics,epsfig,psfig,multirow}
\usepackage{hyperref}
\usepackage{amsmath}
\usepackage{amssymb}	
\usepackage{pdflscape}
\usepackage{footnote}

\newcommand{\rion}[2]{{\ensuremath{\mbox{\rm #1$\,${\small\uppercase\expandafter{\romannumeral#2\relax}}}}}}
\usepackage{newtxtext,newtxmath}
\usepackage[T1]{fontenc}
\usepackage{ae,aecompl}

\usepackage{color}
\usepackage[normalem]{ulem}

\definecolor{darkgreen}{rgb}{0.13, 0.55, 0.13}

\definecolor{brown}{rgb}{0.59, 0.29, 0.0}
\definecolor{ab}{rgb}{0.36, 0.54, 0.66}

\newcommand{\aref}[1]{\hyperref[#1]{Appendix~\ref{#1}}}

\title[Origin of nitrogen]{On the origin of nitrogen at low metallicity}

\author[A. Roy et al.]{Arpita Roy$^{1,2,3}$\thanks{E-mail: arpita.roy1016@gmail.com, arpita.roy@ens-lyon.fr},
Michael A. Dopita$^{2,3\thanks{deceased}}$,
Mark R. Krumholz$^{2,3}$,
Lisa J. Kewley$^{2,3}$,
\newauthor
Ralph S. Sutherland$^{2,3}$,
Alexander Heger$^{3-7}$
\\
$^{1}$\'{E}cole Normale Sup\'{e}rieure de Lyon, Centre de Recherche Astrophysique de Lyon, 46 All\'{e}e d'Italie, 69007 Lyon, France.\\
$^{2}$Research School of Astronomy and Astrophysics, Australian National University, Cotter Road, Weston Creek, ACT 2611, Australia.\\
$^{3}$ARC Centre of Excellence for All Sky Astrophysics in 3 Dimensions (ASTRO 3D), Canberra, ACT 2611, Australia\\
$^{4}$School of Physics and Astronomy, Monash Centre for Astrophysics, 19 Rainforest walk, Monash University, VIC 3800, Australia.\\
$^{5}$Tsung-Dao Lee Institute, Shanghai 200240, China.\\
$^{6}$Australian Research Council Centre of Excellence for Gravitational Wave
Discovery (OzGrav), Clayton, VIC 3800, Australia\\
$^{7}$Joint Institute for
Nuclear Astrophysics – Center for the Evolution of the Elements, 640 S Shaw Lane East Lansing, MI 48824.\\
}

\date{Accepted XXX. Received YYY; in original form ZZZ}

\pubyear{2019}

\begin{document}
\label{firstpage}
\pagerange{\pageref{firstpage}--\pageref{lastpage}}
\maketitle

\begin{abstract}
Understanding the evolution of the N/O ratio in the interstellar medium (ISM) of galaxies is essential if we are to complete our picture of the chemical evolution of galaxies at high redshift, since most observational calibrations of O/H implicitly depend upon the intrinsic N/O ratio. The observed N/O ratio, however, shows large scatter at low O/H, and is strongly dependent on galactic environment. We show that several heretofore unexplained features of the N/O distribution at low O/H can be explained by the N seen in metal-poor galaxies being mostly primary nitrogen that is returned to the ISM via pre-supernova winds from rapidly rotating massive stars ($M \gtrsim 10$ M$_\odot$, $v/v_{\rm crit} \gtrsim 0.4$). This mechanism naturally produces the observed N/O plateau at low O/H. We show that the large scatter in N/O at low O/H also arises naturally from variations in star-formation efficiency. By contrast, models in which the N and O come primarily from supernovae provide a very poor fit to the observed abundance distribution. We propose that the peculiar abundance patterns we observe at low O/H are a signature that dwarf galaxies retain little of their SN ejecta, leaving them with abundance patterns typical of winds.
\end{abstract}



\section{Introduction}
\label{intro_sec}
The chemical history of nitrogen in the Universe is both very important for observations and very poorly understood. Observationally, N is important because accurate measurements of metallicity in high-redshift galaxies are heavily reliant on an accurate estimate of the N/O ratio. For example, one of the best abundance diagnostics is based upon the [\ion{N}{ii}]/[\ion{O}{ii}] ratio \citep{Kewley02}. At high redshift the [\ion{O}{ii}]$\lambda\lambda3727,3729$ doublet is often unobservable, and calibrations based upon the  [\ion{N}{ii}]/H$\alpha$ ratio \citep{Denicolo02} or the [\ion{N}{ii}]/[\ion{O}{iii}] ratio \citep{Pettini04} are used instead. Sometimes, only the red lines of H$\alpha$, [\ion{N}{ii}]$\lambda 6584$ and the [\ion{S}{ii}]$\lambda\lambda6717,6731$ doublet are observed, so the calibration of the O/H ratio must rely on indirect methods of determination using either the  [\ion{N}{ii}]/H$\alpha$ ratio, the  [\ion{N}{ii}]/ [\ion{S}{ii}] ratio, or some judiciously chosen combination of these \citep{Dopita16}. All of these methods rely on us having at least an approximate understanding of the N/O ratio and its evolution with overall metallicity.

However, that understanding is poor, because the chemical origins of nitrogen are complex. Oxygen is a primary element, since its main origin site is in stars undergoing triple-$\alpha$ fusion, a process that does not depend on the presence of a pre-existing supply of C or O in the star that was inherited from the interstellar medium (ISM). Nitrogen, by contrast, can be both a primary and a secondary element, and multiple production sites are possible for both. Secondary production occurs in massive stars undergoing CNO burning catalysed by C inherited from the ISM, either in the core during the main sequence or in a shell after it \citep[e.g.,][]{meynet2002}. Primary production occurs both in intermediate-mass post-main sequence stars undergoing hot bottom burning \citep[e.g.,][]{marigo2001}, and in rapidly-rotating massive stars where CNO burning occurs in a shell, but the C that catalyses the reaction is rotationally dredged up from the helium-burning core rather than having been inherited from the ISM \citep[e.g.,][]{meynet2002}. Which production channel dominates, and under what circumstances, has strong implications for the N/O ratio and its evolution. In simple closed box models, \citet{Edmunds78} showed that the abundances of primary elements scale with the remaining gas fraction $\mu$ as $Z_{\rm P} \propto \ln (1/\mu)$, while those of secondary elements, whose production rate depends on the amount of C and O already created, scale as $Z_{\rm S} \propto [\ln(1/\mu)]^2$. Thus if the predominant N production mechanism is primary, the N/O ratio should be independent of total O metallicity, while if the secondary channel dominates we should expect $\mathrm{N}/\mathrm{O} \propto \mathrm{O}/\mathrm{H}$.

Given the challenges of modelling nitrogen production and the evolution of the N/O ratio theoretically, many  authors have undertaken direct observational studies. There are three primary approaches available. First, one can use metal-poor Milky Way halo stars as archaeological records of N production. Second, one can study N and O abundances in the \ion{H}{ii} regions of the Milky Way and nearby galaxies, with local galaxies at different overall metallicities serving as proxies for studying N production in galaxies at different redshifts. Finally, one can study high-redshift galaxies directly, hoping to observe N and O \textit{in situ}.

Stellar studies suggest that N can be both a primary and a secondary element depending on environment. For example, \citet{israelian2004} survey 31 unevolved halo stars with $-3.1 \leq [\mathrm{Fe}/\mathrm{H}] \leq 0$. They find that $[\mathrm{N}/\mathrm{O}]$ is nearly constant for $[\mathrm{O}/\mathrm{H}] \lesssim -1.8$, suggesting a primary origin for N, but that there is a correlation between $[\mathrm{N}/\mathrm{O}]$ and $[\mathrm{O}/\mathrm{H}]$ for $[\mathrm{O}/\mathrm{H}] \gtrsim -1.8$, suggesting a changeover to secondary N production. \citet{spite2005} analyse 35 stars with $-4 \leq [\mathrm{Fe}/\mathrm{H}] \leq -2$, and confirm this result that there is no systematic trend in $[\mathrm{N}/\mathrm{O}]$ with $[\mathrm{O}/\mathrm{H}]$ at low metallicity; however, they also find that the scatter in  $[\mathrm{N}/\mathrm{O}]$ at low metallicities is very broad, suggesting a more complicated history than simple primary production.

Work on local dwarfs also yields a complex and not easily interpreted set of results. \citet{izotov1999} measure N and O abundances in a sample of 50 low-metallicity ($Z_\odot/50 < Z < Z_\odot/7)$ blue compact galaxies, and find that the N/O ratio is nearly constant for $12+\log(\mathrm{O}/\mathrm{H}) < 7.6$. Other studies of low-metallicity local dwarfs also find that N/O does not systematically vary with O/H at low metallicity, but reveal that there is a $\sim 1$ dex of scatter in the N/O ratio at fixed O/H \citep[e.g.,][]{Kobulnicky96a, Perez-Montero05, Perez-Montero09, vanZee06, Liang06, LopezSanchez10}, similar to the large scatter measured in halo stars. While the lack of systematic N/O variation with O/H might suggest a primary origin for N at low metallicity, even in spite of the large scatter, the local studies also show that N/O ratios can vary substantially on very small ($\lesssim 100$ pc) scales \citep[e.g.,][]{Kobulnicky97, LopezSanchez07, Westmoquette13, Kumari18}. Such local variations will be erased by turbulent mixing on timescales of $\sim 100$ Myr or less \citep[e.g.,][]{Yang12a, Petit15a, Krumholz18a}, so they must be due to a production process with a short timescale, implicating massive stars as the source of the N enhancement. Consistent with this hypothesis, regions of elevated N/O are correlated with both local maxima in the star formation rate and regions where the stellar population shows strong Wolf-Rayet (WR) spectral features, both within a galaxy and over entire galaxies \citep{Brinchmann08}. N enrichment by young, massive stars, suggests a secondary origin, and implies extremely rapid mass return, since both Wolf-Rayet features and the H$\alpha$ emission used to infer star formation rates are indicators of stellar populations $\lesssim 5$ Myr old.

Measurement of N and O \textit{in situ} in high redshift galaxies are challenging for the reasons discussed above. Nonetheless, there are published results from both emission and absorption studies. Measurements that attempt to characterise the N/O ratio via emission from \ion{H}{ii} regions using a variety of methods find that N/O increases with star formation rate \citep{Kojima17}, though \citet{Masters16} argue that stellar mass (which correlated with star formation rate) is in fact the more important variable. However, these high-z galaxies are at much higher metallicities compared to local dwarfs. 
In contrast to the high-redshift emission studies, which mainly target more metal-rich systems, absorption studies in damped Lyman $\alpha$ (DLA) systems probe much lower metallicities. The DLA observations consistently show no correlation between the ratio of N to various $\alpha$ elements and the $\alpha$/H ratio \citep{centurion1998, lu1998, pettini2002, prochaska2002}, similar to the results obtained in \ion{H}{ii} studies in local dwarfs. There is disagreement between the various studies about the amount and pattern of scatter in the N/$\alpha$ ratio, however.

To summarise the observations: studies at low-metallicity, whether in halo stars, local dwarfs, or DLAs, consistently fail to find a correlation between N/O and O/H, favouring a primary origin for N. However, there is extreme scatter in the N/O ratio, suggesting a complex origin mechanism. At higher metallicity, N/O appears to increase with O/H, suggesting a secondary origin to N; N/O also correlates with the total stellar mass and star formation rate. However, both the detailed structure of the relation between N/O and stellar mass or star formation, and the presence of extremely local variations in N/O that are correlated with Wolf-Rayet spectral features, seem to require a channel of N production on short timescales that is associated with massive stars. One might presume this to be a secondary channel, since it is associated with stars undergoing CNO burning.

Our goal in this paper is to revisit the question of the origin of N and its status as a primary or secondary element at low metallicity. In particular, we investigate the hypothesis that the abundance of N at low metallicity, and its relationship to O, can be explained by mass return from massive stars \textit{prior to the onset of supernovae}, as hinted at by the observations showing a strong correlation between N/O and Wolf-Rayet spectral features. To this end, we carry out an extensive grid of stellar evolution calculations, following massive stars across a broad range of rotation rate and metallicity, and we consider a range of scenarios for what might be retained by low-mass, low-metallicity galaxies versus lost to the intergalactic or circumgalactic medium. This paper is organised as follows: in \autoref{method_sec}, we describe our stellar models and theoretical methods to obtain yields and wind velocities; in \autoref{results_sec}, we discuss the results; in \autoref{disc_sec}, we discuss the possible scenarios for the origin of N in the very early universe and summarise our primary findings. 

\section{Methods}
\label{method_sec}
\subsection{Stellar models}
\label{model_sec}

Since we are interested in chemical enrichment via stellar winds prior to supernovae (SNe), we require set of stellar evolution models capable of predicting enrichment rates. For this purpose,
we use MESA \citep{paxton2011, paxton2013, paxton2015} to calculate the evolution of a grid of stellar models described below. Our set of parameter choices in MESA are identical to that of the MESA Isochrone Stellar Tracks (MIST) set of MESA runs described in \citet{choi2016}, with small modifications to improve the treatment of massive stars as described in \citet{roy2019}. The modifications we describe here will also be included in the MESA Isochrone Stellar Tracks-II (MIST-II) models of \citet{dotter2020}. We briefly summarise our setup here for reader convenience, and refer readers to \citet{roy2019} for complete details and exploration of the uncertainties in various parameter choices and their effects:

\begin{itemize}
\item {\it Mixing mechanisms}:  We include the following non-rotational mixing mechanisms: convection, overshoot-convection, and semiconvective mixing. The rotational and magnetic mixing mechanisms we include are dynamical shear instability, Solberg-Hoiland instability, secular shear instability, Eddington-Sweet  circulation, Goldreich-Schubert-Fricke instability and Spruit Torques (see \citet{paxton2013}, \citet{roy2019} and references therein).  We calculate the diffusion coefficients for chemical transport ($D$) and angular momentum transport ($\nu$) due to these mechanisms as described in equations 1 and 2 of \citet{roy2019}. 
\item {\it Wind mass-loss}: We adopt the radiative wind prescriptions known as Dutch mass-loss scheme in MESA. In this scheme, during  evolution on the main sequence, when the effective temperature T$_{\rm{eff}}>10^4$ K and surface hydrogen mass-fraction X$_{\rm{surf}}>0.4$, we use the mass-loss prescription of \citet{vink2001}. During the Wolf-Rayet phase, when the effective temperature T$_{\rm{eff}}>10^4$ K and surface hydrogen mass-fraction X$_{\rm{surf}}<0.4$, we use the \citet{nugis2000} mass-loss prescription. For the cool stars (T$_{\rm{eff}}<10^4$ K), we adopt the \citet{deJager1988} empirical formula. For the detailed justifications of these choices and further explanation of these mass-loss prescriptions in massive stars, see \citet{choi2016}.
\item {\it Initial abundances}: Rather than simply adopting scaled Solar abundances for calculations with $\mathrm{[Fe/H]} \neq 0$, we vary the $\alpha$ abundance [$\alpha$/Fe] as a function of [Fe/H] following the observed empirical scaling shown in figure 2 of \citet{nicholls2017}. Following that paper, we adopt $[\alpha/\mathrm{Fe}] =0.4$ for the iron metallicities $\mathrm{[Fe/H]}\le -1.0$  that we explore in this work.
\end{itemize}    
We use this setup to calculate the evolution of a grid of models with initial mass of $10-150$ $M_\odot$ in steps of $5$ $M_\odot$, at iron metallicities $\mathrm{[Fe/H]} = -4$ to $-2$ in steps of $-1$, and at initial stellar rotation rates from $v/v_{\rm crit} = 0 - 0.6$ in steps of 0.2. We take $v/v_{\rm crit} = 0.4$ to be our fiducial choice because theoretical models of massive star formation suggest that rotation rates in this range should be the norm, independent of metallicity \citep{rosen2012}, and unless otherwise noted all the results we show below are for this case. We run almost all models to the end of core carbon ($^{12}\mathrm{C}$) burning (defined as when the central $^{12}$C mass fraction falls below $10^{-4}$), and a few models to the end O burning (defined as when the central $^{16}$O mass fraction falls below $10^{-4}$). For a few cases the time step becomes so small that it is impractical to continue to this point, and we instead halt either when the central $^{12}$C mass fraction falls below 5\%, or at the end of He burning (when central $^{4}$He is below $10^{-4}$ by mass). However, these cases constitute a small minority of our models -- see detailed discussion in \autoref{model_grid_evol_sec}.

\subsection{Abundance calculation}
\label{abund_sec}
The two primary outputs of our grid of stellar evolution models is a set of time-dependent wind mass loss rates $\dot{M}_w(X, m, t)$, defined as the rate at which a star of initial mass $m$ and age $t$ ejects element $X$ in a wind, and stellar lifetimes $t_\ell(m)$, which we take to be identical to the time to reach the end of C burning (since the remaining lifetime thereafter is small). We supplement these outputs of our evolution calculations with a set of supernova yields $M_{\rm SN}(X,m)$, 
where $M_{\rm SN}(X,m)$ is defined as the mass of element $X$ ejected when a star of initial mass $m$ explodes as a supernova after its lifetime $t_\ell$. We discuss our choice of SN yields in \autoref{ssec:SNe}. The quantities $\dot{M}_w$, $t_\ell$, and $M_{\rm SN}$ are all functions of the initial composition and rotation rate as well, but we do not write out these dependences explicitly for reasons of compactness. 

We consider a simple stellar population formed at time $t=0$ with an initial mass function (IMF) $dn/dm$, normalised to have unit integral. For all the numerical results we present in this paper, we use a \citet{salpeter1955} IMF, with a maximum mass of 150 M$_\odot$. The total mass of element $X$ returned to the gas phase per unit stellar mass (at formation) by the stellar population by the time it reaches age $t$ is
\begin{eqnarray}
\psi_{\rm ret}(X,t) &=& \frac{1}{\langle m \rangle} \int_0^\infty \left[ M_{w}(X,m,t) + \right.
\nonumber \\
& & \left. M_{\rm SN}(X,m) H(t - t_\ell(m))\right] \frac{dn}{dm}
\, dm,
\label{eq:return}
\end{eqnarray}
where
\begin{equation}
\langle m\rangle = \int m\frac{dn}{dm}\,dm
\end{equation}
is the mean stellar mass,
\begin{equation}
M_w(X,m,t) = \int_0^t \dot{M}_w(X,m,t')\, dt'
\end{equation}
is the cumulative mass of element $X$ ejected in winds by a star of initial mass $m$ up to age $t$, and $H(x)$ is the Heaviside step function, which is unity for $x > 0$ and 0 for $x< 0$. In \autoref{eq:return}, the first term in the integral represents the contribution from winds, and the second represents the contribution from supernovae. We can analogously define the mass returns $\psi_w$ and $\psi_{\rm SN}$ from winds and supernovae alone. If the lifetime $t_\ell(m)$ is a monotonically-decreasing function of $m$, as is the case for our model grid, then this expression can also be simplified to

\begin{eqnarray}
\lefteqn{\psi_{\rm ret}(X,t) = \psi_w(X,t) + \psi_{\rm SN}(X,t) =}
\label{eq:psi_Xt}
\\
& & \frac{1}{\langle m \rangle} \left[
\int_0^\infty M_{w}(X,m,t) \frac{{\mathrm{d}}n}{{\mathrm{d}}m} \, {\mathrm{d}}m +
\int_{m_{\rm d}}^\infty M_{\rm SN}(X,m) \frac{{\mathrm{d}}n}{{\mathrm{d}}m} \, {\mathrm{d}}m\right],
\nonumber
\end{eqnarray}

where $m_{\rm d}$ is the ``death mass'' at age $t$, given implicitly by $t_\ell(m_{\rm d}) = t$.

Now consider a gaseous reservoir of mass $M_{\rm g}$, which instantaneously converts a fraction $\epsilon_*$ of its mass to stars. The mass fractions of element $X$ in the gas prior to this event are $f_0(X)$. A time $t$ after the star formation event, the mass of any element $X$ in the gas phase is
\begin{equation}
M(X,t) = \left[(1 - \epsilon_*) f_0(X) + \epsilon_* \psi_{\rm ret}(X,t)\right] M_{\rm g}.
\label{eq:M_Xt}
\end{equation}
Here the first term represents the mass of element $X$ left in the gas phase after star formation, while the second term represents the amount returned by stellar evolution. We can therefore write the abundance ratio of any two elements as a function of time as: 
\begin{equation}
\frac{M(X,t)}{M(Y,t)} = \frac{(1 - \epsilon_*) f_{0}(X) + \epsilon_* \psi_{\rm ret}(X,t)}{(1 - \epsilon_*) f_{0}(Y) + \epsilon_* \psi_{\rm ret}(Y,t)}\, .
\label{eq:Xi_Xj}
\end{equation}
We can also consider the effects of adding additional primordial gas to this reservoir. Let $f_{\rm p}(X)$ be the fractional abundance of element $X$ in this primordial gas, and let $M_{\rm p}(t)$ be the amount of primordial gas mixed into the reservoir by time $t$. In this case the abundance ratio generalises to
\begin{equation}
\frac{M(X,t)}{M(Y,t)} = \frac{(1 - \epsilon_*) f_{0}(X) + \epsilon_* \psi_{\rm ret}(X,t) + \epsilon_{\rm p}(t) f_{\rm p}(X)}{(1 - \epsilon_*) f_{0}(Y) + \epsilon_* \psi_{\rm ret}(Y,t) + \epsilon_{\rm p}(t) f_{\rm p}(Y)} \, ,
\label{eq:Xi_Xj_pri}
\end{equation}
where $\epsilon_{\rm p}(t) = M_{\rm p}(t) / M_{\rm g}$ is the fractional mass of primordial gas added. Again, we can straightforwardly define analogous quantities for yields due to wind or SNe alone. In terms of number fractions, the abundance ratio can be written as:
\begin{equation}
\frac{N(X,t)}{N(Y,t)} = \frac{M(X,t)}{M(Y,t)} \frac{m_Y}{m_X} \, ,
\label{eq:Ni_Nj}
\end{equation}
where $m_X$ is the atomic mass of element $X$.  We will denote the ratio $N(X)/N(Y)$ by the usual shorthand $X/Y$ in the rest of the paper.

\subsection{Stellar wind velocity}
\label{vwind_sec}

In order to address the question of whether wind ejecta are likely to be retained by the galaxies into which they are injected, we wish to compute wind velocities as well as yields. Because calculation of wind velocities is not included in the  standard MESA outputs,\footnote{MESA does in fact output a quantity that it describes as the wind velocity, but this is a rough estimate of the characteristic speed at the base of the photosphere, not an estimate of the (generally much larger) terminal velocities produced by line-driven acceleration of massive star winds.} we post-process the MESA outputs to generate this estimate.
To calculate the wind velocity ($v_{\rm wind}$) of O stars, we 
adopt the same wind models as those used for mass loss, with some modifications as described below. 
For O stars, which we characterise as those with $T_{\rm eff} > 1.1\times 10^4$ K and surface hydrogen abundance $X_{\rm H} > 0.4$, we use the \citet{vink2000, vink2001} prescription whereby the wind velocity taken to be $1.3  v_{\rm esc} Z_{\rm surf}^{0.13}$ for stars on the cool side of the bistability jump, and $2.6 v_{\rm esc} Z_{\rm surf}^{0.13}$ for stars on the hot side of the bistability jump; here $Z_{\rm surf}$ is the surface metallicity, and $v_{\rm esc}$ is the surface escape speed. We determine whether a given star is on the hot or cool side of the bistability jump from \citeauthor{vink2001}'s equation 15.

For Wolf-Rayet (WR) stars, which are characterised by $T_{\rm eff} > 10^4$ but $X_{\rm H} < 0.4$, our mass loss prescription comes from \citet{nugis2000}. However, we cannot directly adopt their recommended empirical scaling for wind speed with stellar parameters, because their scaling is calibrated entirely at metallicities much higher than the values with which we are concerned, and a simple extrapolation of their scalings to the metallicity range relevant to our calculations leads to unphysically-large velocities (in some cases exceeding several percent of $c$). Instead, we note that, since WR winds are radiatively-driven, the natural momentum scale for their winds is $L/c$, where $L$ is the stellar luminosity. \citet{nugis2000} parameterise this relationship by
\begin{equation}
v_{\rm wind} = \eta \frac{L}{\dot{M} c}, 
\label{eq:v_wind}
\end{equation}
and find that $\eta \approx 1$ for all their observed stars. We therefore model WR star wind velocities using \autoref{eq:v_wind} with $\eta = 1$.

Finally, for stars whose effective temperatures are below $10^4$ K, our mass loss prescription is taken from the empirical fits of \citet{deJager1988}, but these authors do not provide an estimate of wind speeds. We therefore adopt the dust-driven wind model scaling of \citet{elitzur2001}, $v_{\rm wind} \propto R_{\rm gd}^{-1/2} L^{1/4}$, where $R_{\rm gd}$ is the gas-to-dust ratio. We set the normalisation using the empirical calibration of \citet{goldman17a}, who find a wind speed of $9.4$ km s$^{-1}$ for a Milky Way gas-to-dust ratio; we assume that the gas-to-dust ratio scales inversely with the stellar surface metallicity $Z_{\rm surf}$. For stars at intermediate temperatures, $T = 1.0 - 1.1\times 10^4$ K, we linearly interpolate between this case and the O star case described above.


Given the above prescriptions for the wind velocity of each star, we calculate the mass loss weighted wind velocity as: 
\begin{equation}
\langle {v}_{\rm wind}\rangle _{\dot{M}} ={ {{\int_{M_{i}}^{M_{f}}} {v}_{\rm wind} \dot{M}\frac{{\mathrm{d}}n}{{\mathrm{d}}m} \, {\mathrm{d}}m} \over {\int_{M_{i}}^{M_{f}} \dot{M}\frac{{\mathrm{d}}n}{{\mathrm{d}}m} \, {\mathrm{d}}m} }\, ,
\label{eq:vwind}
\end{equation}
where $M_{\mathrm i}$ and $M_{\mathrm f}$ are our minimum (10 M$_\odot$) and maximum (150 M$_\odot$) mass from our grid as described in \autoref{model_sec}.


\subsection{SN Models}
\label{ssec:SNe}

In order to compare the yields and ejection velocities we compute from stellar winds to those produced in SN explosions, we require a set of SN models. Any such choice is complicated by the large uncertainties in the ultimate fates of massive stars, with possibilities that include not only ordinary Type II/Ib/Ic SNe, but also direct collapse to black hole, pulsational pair-instability SNe (PPISN), and ordinary pair instability SNe (PISN). The ranges of initial mass, rotation rate, and metallicity that map to these possible outcomes are substantially uncertain, and likely non-monotonic in one or more of the variables \citep[e.g.,][]{sukhbold16}. Given the large uncertainties, we adopt a conservative approach and consider only ordinary Type II SNe, which is almost certainly the fate of all massive stars with initial masses $m\lesssim 15-20$ $M_\odot$. More massive stars, depending on their CO core mass at the end of their evolution, may explode as PISN, and in that case we will have higher SNe ejecta velocities and yields compared to our results presented in this paper. Our results, therefore, provide the lower limits of those quantities.  

For stars that end their lives as ordinary Type II SNe, we take our SN yields $M_{\rm SN}(X,m)$ from the published tables of \citet{limongi2018}. We match our models to \citeauthor{limongi2018}'s tables based on the He core mass, i.e., for each of our models we measure the He core mass $M_{\rm He,core}$ at the end of our MESA simulation, and we interpolate \citeauthor{limongi2018}'s tabulated yields as a function of $M_{\rm He,core}$ to set the yield for our model. Note that the interpolation based on the CO core mass, $M_{\rm CO,core}$, also produces the same results. Their metallicity grid is only partly overlapping with ours -- they provide yields for $[\mathrm{Fe}/\mathrm{H}] = -3$ and $-2$, but not $-4$, and thus we will omit comparisons between winds and SN for our $[\mathrm{Fe}/\mathrm{H}] = -4$ case. We use \citeauthor{limongi2018}'s models with rotation speed $v = 300$ km s$^{-1}$, since this is close to our fiducial choice $v/v_{\rm crit} = 0.4$. Finally, we set $M_{\rm SN}(X,m) = 0$ for all stars with $m>15$ M$_\odot$, implicitly assuming that these stars collapse directly to black holes because the mass limit below which stars successfully explode as SNe has an uncertainty between 15-20 M$_\odot$. We discuss the dependence of our results on this choice in \aref{upper_mass_sec}.

In addition to the yields, we estimate SN ejecta velocities. The mean ejection velocity is
\begin{equation}
\langle {v}_{\rm SN}\rangle _{M} ={ {\int_{M_{i}}^{M_{f}} \left(\sqrt{2 E_{\rm SN}/m_{\rm {ej}}}\right) {m_{\rm {ej}}\frac{{\mathrm{d}}n}{{\mathrm{d}}m} \, {\mathrm{d}}m}} \over {\int_{M_{i}}^{M_{f}} {m_{\rm {ej}}}\frac{{\mathrm{d}}n}{{\mathrm{d}}m} \, {\mathrm{d}}m} },
\label{eq:vSN}
\end{equation}
where $E_{\rm SN} = 10^{51}$ erg is the SN explosion energy, $m_{\rm {ej}}$ is the SN-ejecta mass. We assume stars that explode successfully as SN leave behind the neutron stars of mass $1.4 \, \mathrm{M}_\odot$ and thus $m_{\rm {ej}} = m_{\rm PSN} - 1.4$ M$\odot$, where $m_{PSN}$ is the pre-SN (PSN) mass (implicitly a function of $m$), $M_{\mathrm i} = 10$ $M_\odot$ and $M_{\mathrm f} = 15$ $M_\odot$. We consider the PSN-mass as the mass at the end of core C burning because after that stars lose hardly any mass until the PSN phase. We find that our PSN masses are similar to \citet{limongi2018} (see \autoref{tab:WR_feh_m2_tab}, \autoref{tab:WR_feh_m3_tab}, \autoref{tab:WR_feh_m4_tab}). The detailed stellar data at the end of their evolutions for our model-grids are presented in \autoref{model_grid_evol_sec}.

\subsection{Velocity dependent yield calculation}
\label{ssec:psi_v}
In order to estimate the velocity with which each element is ejected, we define the velocity-dependent cumulative yield up to time $t$ for velocity $v$, similar to \autoref{eq:psi_Xt}, as:
\begin{eqnarray}
\psi_{\rm ret}(X,v,t) &=& \frac{1}{\langle m \rangle} \left [\int_0^\infty M_{w}(X,m,v,t) \frac{dn}{dm}\, dm + \right. \nonumber\\
& & \left. \int_{m_{\rm d}}^\infty M_{\rm SN}(X,m,v) H(v_{\rm SN} - v) \frac{dn}{dm}\, dm \right ] \, ,
\label{eq:psi_v}
\end{eqnarray}
where $M_{w}(X,m,v,t) = \int_0^t \dot{M}_{w}(X,m,v,t') H(v_w-v) dt'$, and $H(x)$ is the Heaviside step function with $H(x) = 0$ for $x < 0$ and $H(x) = 1$ for $x > 0$. The quantity $\psi_{\rm ret}(X,v,t)$ is simply the mass return of element $X$ up to time $t$, \textit{counting only material that is ejected at velocity $\leq v$}. This can serve as a rough proxy for the yield of elements that are likely to be retained in a galaxy with a low escape speed, rather than escaping into the circumgalactic medium (CGM). For convenience we will also define $\psi_{\rm ret}(X,v)$ as $\psi_{\rm ret}(X,v,t)$ evaluated with $t$ equal to the time at which the last SN occurs, i.e., $\psi_{\rm ret}(X,v)$ is the velocity-dependent yield of all the material produced by stars massive enough to end their lives either in a supernova or as a black hole, and we will define $X/Y(v)$ as the abundance ratio of this material, evaluated by using $\psi_{\rm ret}(X,v)$ in \autoref{eq:Ni_Nj}.
In addition to the velocity-dependent cumulative yields, we can also define velocity-dependent differential yields as ${\rm d}\psi_{\rm ret}/{\rm d}\log v$. Integrated over a bin of finite size $\Delta \log v$, the differential yield in that velocity bin is defined as: 
\begin{equation}
\frac{\mathrm{d}\psi_{\rm ret} (X,v)}{\mathrm{d}\log(v)}  = \frac{\psi_{\rm ret}\left(X, 10^{\log v+\Delta v/2}\right) - \psi_{\rm ret}\left(X, 10^{\log v-\Delta v/2}\right)}{\Delta \log v}.
\label{eq:diffPsi_dvbin}
\end{equation}
As for time-dependent yields, we can also separately define $\psi_{\rm w}(X,v)$ and $\psi_{\rm SN}(X,v)$, and their differential equivalents, as the yields due to winds and supernovae alone.



\section{Results}
\label{results_sec}

Here we describe the results of the stellar evolution models, first examining the model grids in \autoref{model_grid_evol_sec}, yields from winds and SNe, and the nucleosynthetic origin of N in massive stars in \autoref{time_yield_sec}, the velocity of the ejected material in \autoref{ssec:velocity}, and then exploring predicted values of N/O and O/H in various scenarios in \autoref{yields_comp_sec}. We explore how the N/O ratio depends on stellar rotation rates in \autoref{rot_dep_sec}.

\begin{landscape}
\begin{table}
\centering
\begin{tabular}{ccccccccccccccc}
\hline\hline
$M_{\rm{initial}}$ & 
Final &
$t_{\rm halt}$ &
$M_{\rm final}$ &
$M_{\rm He,core}$ &
$M_{\rm CO,core}$ &
$\log L$ &
$\log T_{\rm eff}$ &
$v_{\rm crit,eq}$ &
$v/v_{\rm crit,eq}$ &
$X_{\rm H,surf}$ &
$X_{\rm He,surf}$ &
$X_{\rm C,surf}$ &
$X_{\rm N,surf}$ &
$X_{\rm O,surf}$ \\
(M$_\odot$) &
burning &
(Myr) &
(M$_\odot$) &
(M$_\odot$) &
(M$_\odot$) &
(L$_\odot$) &
(K) &
(km s$^{-1}$) \\
\hline
10 & O$_{\rm dep}$ & 3.65e+01 & 8.61e+00 & 7.19e+00 & 5.28e+00 & 1.91e+00 & 3.67e+00 & 4.95e+01 & 2.18e-04 & 5.97e-01 & 4.03e-01 & 3.16e-06 & 9.08e-05 & 7.72e-05 \\
15 & C$_{\rm dep}$ & 1.47e+01 & 1.48e+01 & 5.17e+00 & 3.23e+00 & 5.07e+00 & 3.63e+00 & 6.64e+01 & 1.43e-02 & 6.87e-01 & 3.13e-01 & 2.23e-06 & 8.89e-05 & 8.09e-05 \\
20 & C$_{\rm dep}$ & 1.06e+01 & 1.90e+01 & 7.42e+00 & 4.88e+00 & 5.29e+00 & 3.62e+00 & 6.56e+01 & 2.75e-03 & 6.81e-01 & 3.19e-01 & 1.89e-06 & 9.19e-05 & 7.81e-05 \\
25 & C$_{\rm dep}$ & 9.01e+00 & 2.38e+01 & 9.67e+00 & 7.24e+00 & 5.53e+00 & 3.64e+00 & 6.60e+01 & 7.15e-03 & 6.48e-01 & 3.52e-01 & 1.56e-06 & 1.09e-04 & 5.98e-05 \\
30 & C$_{\rm dep}$ & 8.85e+00 & 2.66e+01 & 2.05e+01 & 1.01e+01 & 5.95e+00 & 4.13e+00 & 1.13e+02 & 1.26e-02 & 3.24e-01 & 6.76e-01 & 1.11e-06 & 1.45e-04 & 1.87e-05 \\
35 & C$_{\rm dep}$ & 7.87e+00 & 3.12e+01 & 2.54e+01 & 1.14e+01 & 6.09e+00 & 4.29e+00 & 1.64e+02 & 1.06e-02 & 2.70e-01 & 7.29e-01 & 1.09e-06 & 1.52e-04 & 1.10e-05 \\
40 & C$_{\rm dep}$ & 6.73e+00 & 3.64e+01 & 2.65e+01 & 1.35e+01 & 6.07e+00 & 4.18e+00 & 1.35e+02 & 8.77e-03 & 3.12e-01 & 6.88e-01 & 1.36e-06 & 1.53e-04 & 9.74e-06 \\
45 & C$_{\rm dep, \, marginal}$ & 6.04e+00 & 4.11e+01 & 2.85e+01 & 1.66e+01 & 6.11e+00 & 4.54e+00 & 2.62e+02 & 1.02e-01 & 3.08e-01 & 6.92e-01 & 1.49e-06 & 1.56e-04 & 5.91e-06 \\
50 & C$_{\rm dep, \, marginal}$ & 5.60e+00 & 4.35e+01 & 3.28e+01 & 2.04e+01 & 6.24e+00 & 4.47e+00 & 2.65e+02 & 2.13e-02 & 2.12e-01 & 7.87e-01 & 2.22e-06 & 1.58e-04 & 1.80e-06 \\
55 & C$_{\rm dep}$ & 5.31e+00 & 3.11e+01 & 3.11e+01 & 2.61e+01 & 6.20e+00 & 5.45e+00 & 1.26e+03 & 6.42e-01 & 7.76e-29 & 1.56e-01 & 4.38e-01 & 9.67e-09 & 4.05e-01 \\
60 & C$_{\rm dep}$ & 5.01e+00 & 2.89e+01 & 2.89e+01 & 2.41e+01 & 6.16e+00 & 5.44e+00 & 1.21e+03 & 6.29e-01 & 1.16e-24 & 1.49e-01 & 4.09e-01 & 1.16e-13 & 4.41e-01 \\
65 & C$_{\rm dep}$ & 4.77e+00 & 2.75e+01 & 2.75e+01 & 2.25e+01 & 6.13e+00 & 5.44e+00 & 1.18e+03 & 5.51e-01 & 7.20e-26 & 1.76e-01 & 4.20e-01 & 3.55e-09 & 4.03e-01 \\
70 & C$_{\rm dep}$ & 4.56e+00 & 2.80e+01 & 2.80e+01 & 2.34e+01 & 6.14e+00 & 5.44e+00 & 1.16e+03 & 5.12e-01 & 2.91e-25 & 1.75e-01 & 4.20e-01 & 1.20e-10 & 4.04e-01 \\
75 & C$_{\rm dep}$ & 4.37e+00 & 3.22e+01 & 3.22e+01 & 2.69e+01 & 6.21e+00 & 5.45e+00 & 1.22e+03 & 5.61e-01 & 1.71e-26 & 1.62e-01 & 4.05e-01 & 2.20e-14 & 4.31e-01 \\
80 & C$_{\rm dep}$ & 4.22e+00 & 3.19e+01 & 3.19e+01 & 2.65e+01 & 6.21e+00 & 5.44e+00 & 1.18e+03 & 5.40e-01 & 7.12e-26 & 1.38e-01 & 3.88e-01 & 2.04e-12 & 4.73e-01 \\
85 & C$_{\rm dep}$ & 4.09e+00 & 3.13e+01 & 3.13e+01 & 2.60e+01 & 6.20e+00 & 5.44e+00 & 1.16e+03 & 4.79e-01 & 3.84e-25 & 1.41e-01 & 3.94e-01 & 5.31e-14 & 4.64e-01 \\
90 & C$_{\rm dep}$ & 3.96e+00 & 3.52e+01 & 3.52e+01 & 2.95e+01 & 6.27e+00 & 5.45e+00 & 1.18e+03 & 4.99e-01 & 6.22e-25 & 1.24e-01 & 3.70e-01 & 1.78e-18 & 5.06e-01 \\
95 & C$_{\rm dep}$ & 3.85e+00 & 3.85e+01 & 3.85e+01 & 3.24e+01 & 6.31e+00 & 5.45e+00 & 1.23e+03 & 5.61e-01 & 2.11e-25 & 1.07e-01 & 3.39e-01 & 1.46e-17 & 5.53e-01 \\
100 & C$_{\rm dep}$ & 3.76e+00 & 3.93e+01 & 3.93e+01 & 3.30e+01 & 6.32e+00 & 5.45e+00 & 1.22e+03 & 5.11e-01 & 2.22e-26 & 1.36e-01 & 3.71e-01 & 4.54e-18 & 4.92e-01 \\
105 & C$_{\rm dep}$ & 3.64e+00 & 5.52e+01 & 5.52e+01 & 4.74e+01 & 6.49e+00 & 5.45e+00 & 1.38e+03 & 6.86e-01 & 6.36e-25 & 1.09e-01 & 3.06e-01 & 1.14e-17 & 5.84e-01 \\
110 & C$_{\rm dep}$ & 3.58e+00 & 4.53e+01 & 4.53e+01 & 3.90e+01 & 6.40e+00 & 5.45e+00 & 1.24e+03 & 5.42e-01 & 1.59e-24 & 1.04e-01 & 3.18e-01 & 5.64e-19 & 5.75e-01 \\
115 & C$_{\rm dep}$ & 3.49e+00 & 5.44e+01 & 5.44e+01 & 4.62e+01 & 6.49e+00 & 5.46e+00 & 1.33e+03 & 5.85e-01 & 1.10e-24 & 1.03e-01 & 2.98e-01 & 3.90e-19 & 5.96e-01 \\
120 & C$_{\rm dep}$ & 3.43e+00 & 5.44e+01 & 5.44e+01 & 4.67e+01 & 6.49e+00 & 5.45e+00 & 1.28e+03 & 5.63e-01 & 2.56e-23 & 1.13e-01 & 3.12e-01 & 4.73e-19 & 5.73e-01 \\
125 & C$_{\rm dep}$ & 3.35e+00 & 9.09e+01 & 8.97e+01 & 8.64e+01 & 6.79e+00 & 5.41e+00 & 1.65e+03 & 1.13e-01 & 7.13e-03 & 9.93e-01 & 1.29e-06 & 1.61e-04 & 7.33e-07 \\
130 & C$_{\rm dep}$ & 3.29e+00 & 9.05e+01 & 9.02e+01 & 7.79e+01 & 6.83e+00 & 5.47e+00 & 1.89e+03 & 1.18e-01 & 1.42e-04 & 1.00e+00 & 1.66e-06 & 1.60e-04 & 5.70e-07 \\
135 & C$_{\rm dep}$ & 3.24e+00 & 9.72e+01 & 9.52e+01 & 8.97e+01 & 6.84e+00 & 5.41e+00 & 1.63e+03 & 6.92e-02 & 2.52e-02 & 9.75e-01 & 1.19e-06 & 1.61e-04 & 7.62e-07 \\
140 & C$_{\rm dep, \, marginal}$ & 3.21e+00 & 8.76e+01 & 8.76e+01 & 5.09e+01 & 8.17e+00 & 5.82e+00 & 4.20e+01 & 1.99e+01 & 1.84e-25 & 9.08e-02 & 2.39e-01 & 2.08e-16 & 6.66e-01 \\
145 & C$_{\rm dep, \, marginal}$ & 3.18e+00 & 9.74e+01 & 9.74e+01 & 6.09e+01 & 9.84e+00 & 6.21e+00 & 4.19e+01 & 1.26e+01 & 3.41e-24 & 1.30e-01 & 2.92e-01 & 1.27e-12 & 5.75e-01 \\
150 & C$_{\rm dep, \, marginal}$ & 3.14e+00 & 1.05e+02 & 1.05e+02 & 5.77e+01 & 7.19e+00 & 5.57e+00 & 1.75e+03 & 2.01e-01 & 3.36e-08 & 1.00e+00 & 1.41e-04 & 2.08e-04 & 3.28e-05 \\
\hline\hline
\end{tabular}
\caption{
\label{tab:WR_feh_m2_tab}
Stellar properties at the final time reached in our MESA simulations for [Fe/H]=-2.0 ([$\alpha$/Fe]=0.4),  $v/v_{\rm crit}=0.4$. Column meanings are as follows: (1) initial mass of the star, (2) chemical state of the core at the end of the simulation -- can be either O$_{\rm {dep}}$ (oxygen-depleted: simulation halted when $^{16}{\mathrm O}$ mass fraction drops to 10$^{-4}$), C$_{\rm dep}$ (simulation halted when $^{12}{\mathrm C}$ mass fraction $\leq 10^{-4}$), C$_{\rm {dep, \, marginal}}$ (simulation halted when $^{12}{\mathrm C}$ mass fraction $\leq 0.05$), or He$_{\rm dep}$ (simulation halted when $^{4}{\mathrm{He}}$ mass fraction $< 10^{-4}$), (3) time at which simulation was halted, (4) final mass of the star, (5) He-core mass (the outermost region where $^{1}{\mathrm H}$ mass fraction drops to 10$^{-4}$), (6) CO-core mass (the outermost region where $^{4}{\mathrm {He}}$ mass fraction drops to 10$^{-4}$; note that our results are robust and do not change for the choice of $^{4}{\mathrm {He}}$ mass fraction of 0.01), (7) log surface luminosity, (8) log effective temperature, (9) critical linear velocity at the equator, (10) $v/v_{\rm crit}$ at the equator, (11) surface H mass fraction, (12) surface $^{4}\mathrm{He}$ mass fraction, (13) surface $^{12}\mathrm{C}$ mass fraction, (14) surface $^{14}\mathrm{N}$ mass fraction, (15) surface $^{16}\mathrm{O}$ mass fraction.
}
\end{table}
\end{landscape}

\begin{landscape}
\begin{table}
\centering
\begin{tabular}{ccccccccccccccc}
\hline\hline
$M_{\rm{initial}}$ & 
Final &
$t_{\rm halt}$ &
$M_{\rm final}$ &
$M_{\rm He,core}$ &
$M_{\rm CO,core}$ &
$\log L$ &
$\log T_{\rm eff}$ &
$v_{\rm crit,eq}$ &
$v/v_{\rm crit,eq}$ &
$X_{\rm H,surf}$ &
$X_{\rm He,surf}$ &
$X_{\rm C,surf}$ &
$X_{\rm N,surf}$ &
$X_{\rm O,surf}$ \\
(M$_\odot$) &
burning &
(Myr) &
(M$_\odot$) &
(M$_\odot$) &
(M$_\odot$) &
(L$_\odot$) &
(K) &
(km s$^{-1}$) \\
\hline
10 & C$_{\rm dep}$ & 2.62e+01 & 9.96e+00 & 3.15e+00 & 1.90e+00 & 4.77e+00 & 3.65e+00 & 6.76e+01 & 2.04e-02 & 7.19e-01 & 2.81e-01 & 1.73e-07 & 9.42e-06 & 7.56e-06 \\
15 & C$_{\rm dep}$ & 1.46e+01 & 1.49e+01 & 5.11e+00 & 3.26e+00 & 5.08e+00 & 3.65e+00 & 6.92e+01 & 1.87e-02 & 6.98e-01 & 3.02e-01 & 1.21e-07 & 1.03e-05 & 6.69e-06 \\
20 & C$_{\rm dep}$ & 1.05e+01 & 1.98e+01 & 6.65e+00 & 4.27e+00 & 5.25e+00 & 4.13e+00 & 1.61e+02 & 8.06e-01 & 7.16e-01 & 2.84e-01 & 1.25e-07 & 9.74e-06 & 7.30e-06 \\
25 & C$_{\rm dep}$ & 8.66e+00 & 2.45e+01 & 8.79e+00 & 5.91e+00 & 5.50e+00 & 4.12e+00 & 1.52e+02 & 9.50e-01 & 6.77e-01 & 3.23e-01 & 8.92e-08 & 1.08e-05 & 6.13e-06 \\
30 & C$_{\rm dep, \, marginal}$ & 8.59e+00 & 2.81e+01 & 1.54e+01 & 1.26e+01 & 6.22e+00 & 3.85e+00 & 5.86e+01 & 3.48e-03 & 5.10e-01 & 4.90e-01 & 9.45e-08 & 1.34e-05 & 3.14e-06 \\
35 & C$_{\rm dep}$ & 7.45e+00 & 3.25e+01 & 2.08e+01 & 1.60e+01 & 5.97e+00 & 3.94e+00 & 7.69e+01 & 5.78e-03 & 4.91e-01 & 5.09e-01 & 9.31e-08 & 1.40e-05 & 2.46e-06 \\
40 & C$_{\rm dep}$ & 6.82e+00 & 3.77e+01 & 2.73e+01 & 2.16e+01 & 6.08e+00 & 4.16e+00 & 1.31e+02 & 1.57e-01 & 3.51e-01 & 6.49e-01 & 1.09e-07 & 1.53e-05 & 1.25e-06 \\
45 & C$_{\rm dep}$ & 6.07e+00 & 4.22e+01 & 2.92e+01 & 2.44e+01 & 6.11e+00 & 4.15e+00 & 1.32e+02 & 1.10e-01 & 3.89e-01 & 6.11e-01 & 9.55e-08 & 1.53e-05 & 1.09e-06 \\
50 & C$_{\rm dep, \, marginal}$ & 5.55e+00 & 4.54e+01 & 3.21e+01 & 1.25e+01 & 6.19e+00 & 4.30e+00 & 1.90e+02 & 1.22e-01 & 3.59e-01 & 6.41e-01 & 1.04e-07 & 1.56e-05 & 6.24e-07 \\
55 & C$_{\rm dep}$ & 5.31e+00 & 4.13e+01 & 4.13e+01 & 3.41e+01 & 6.27e+00 & 5.40e+00 & 1.88e+03 & 1.26e+00 & 5.72e-25 & 7.76e-02 & 3.13e-01 & 6.86e-08 & 6.08e-01 \\
60 & C$_{\rm dep}$ & 4.98e+00 & 4.33e+01 & 4.33e+01 & 3.68e+01 & 6.34e+00 & 5.42e+00 & 1.81e+03 & 1.01e+00 & 2.96e-25 & 9.89e-02 & 3.34e-01 & 2.96e-08 & 5.67e-01 \\
65 & C$_{\rm dep}$ & 4.72e+00 & 5.28e+01 & 5.11e+01 & 4.45e+01 & 6.45e+00 & 5.21e+00 & 1.41e+03 & 1.50e+00 & 5.29e-02 & 9.47e-01 & 2.62e-07 & 2.96e-05 & 1.46e-07 \\
70 & C$_{\rm dep}$ & 4.52e+00 & 5.23e+01 & 5.23e+01 & 4.41e+01 & 6.43e+00 & 5.41e+00 & 1.98e+03 & 1.16e+00 & 1.77e-23 & 4.92e-02 & 2.26e-01 & 2.79e-06 & 7.20e-01 \\
75 & C$_{\rm dep}$ & 4.34e+00 & 4.84e+01 & 4.84e+01 & 4.00e+01 & 6.41e+00 & 5.44e+00 & 1.63e+03 & 9.09e-01 & 6.88e-25 & 1.18e-01 & 3.31e-01 & 1.11e-08 & 5.49e-01 \\
80 & C$_{\rm dep}$ & 4.49e+00 & 1.77e+01 & 1.77e+01 & 1.44e+01 & 5.89e+00 & 5.43e+00 & 1.08e+03 & 3.06e-01 & 1.82e-26 & 2.35e-01 & 4.74e-01 & 3.29e-09 & 2.90e-01 \\
85 & C$_{\rm dep}$ & 4.04e+00 & 6.94e+01 & 6.82e+01 & 5.43e+01 & 6.63e+00 & 5.29e+00 & 1.71e+03 & 1.63e+00 & 7.32e-02 & 9.27e-01 & 1.69e-07 & 2.09e-05 & 1.68e-07 \\
90 & C$_{\rm dep}$ & 3.92e+00 & 7.27e+01 & 6.95e+01 & 5.72e+01 & 6.64e+00 & 5.08e+00 & 1.08e+03 & 1.24e+00 & 5.76e-02 & 9.41e-01 & 3.50e-05 & 1.60e-03 & 6.01e-05 \\
95 & C$_{\rm dep}$ & 3.81e+00 & 7.77e+01 & 7.31e+01 & 6.23e+01 & 6.67e+00 & 5.06e+00 & 1.04e+03 & 1.08e+00 & 6.71e-02 & 9.32e-01 & 1.19e-05 & 4.91e-04 & 9.23e-06 \\
100 & C$_{\rm dep}$ & 3.72e+00 & 8.18e+01 & 7.85e+01 & 6.73e+01 & 6.64e+00 & 5.10e+00 & 1.20e+03 & 9.12e-01 & 8.76e-02 & 9.12e-01 & 3.34e-07 & 2.79e-05 & 1.12e-07 \\
105 & C$_{\rm dep, \, marginal}$ & 3.63e+00 & 8.93e+01 & 8.03e+01 & 4.13e+01 & 6.71e+00 & 4.99e+00 & 8.37e+02 & 7.77e-01 & 8.51e-02 & 9.15e-01 & 4.38e-07 & 1.55e-05 & 4.22e-08 \\
110 & C$_{\rm dep}$ & 3.55e+00 & 9.12e+01 & 8.46e+01 & 7.95e+01 & 6.69e+00 & 5.13e+00 & 1.32e+03 & 1.05e+00 & 1.09e-01 & 8.91e-01 & 2.35e-07 & 1.59e-05 & 8.17e-08 \\
115 & C$_{\rm dep, \, marginal}$ & 3.46e+00 & 9.68e+01 & 8.42e+01 & 6.62e+01 & 6.76e+00 & 4.96e+00 & 7.89e+02 & 7.74e-01 & 1.29e-01 & 8.71e-01 & 4.98e-07 & 1.55e-05 & 3.20e-08 \\
120 & C$_{\rm dep}$ & 3.45e+00 & 1.02e+02 & 9.15e+01 & 8.91e+01 & 6.87e+00 & 4.76e+00 & 3.16e+02 & 2.03e-01 & 2.64e-02 & 8.96e-01 & 3.12e-04 & 1.74e-02 & 5.21e-02 \\
125 & C$_{\rm dep}$ & 3.38e+00 & 1.07e+02 & 9.26e+01 & 9.00e+01 & 6.90e+00 & 4.56e+00 & 1.87e+02 & 2.06e-01 & 1.69e-02 & 8.14e-01 & 5.50e-04 & 3.21e-02 & 1.09e-01 \\
130 & C$_{\rm dep, \, marginal}$ & 3.32e+00 & 1.11e+02 & 1.00e+02 & 3.28e+01 & 6.88e+00 & 5.03e+00 & 8.24e+02 & 6.72e-01 & 7.66e-02 & 9.23e-01 & 9.42e-07 & 1.44e-05 & 4.59e-08 \\
135 & C$_{\rm dep}$ & 3.29e+00 & 1.16e+02 & 1.00e+02 & 7.57e+01 & 6.97e+00 & 4.55e+00 & 1.83e+02 & 1.67e-01 & 2.13e-02 & 8.37e-01 & 4.78e-04 & 2.78e-02 & 9.33e-02 \\
140 & C$_{\rm dep, \, marginal}$ & 3.24e+00 & 1.20e+02 & 1.10e+02 & 9.44e+01 & 7.42e+00 & 5.30e+00 & 2.30e+01 & 4.44e+01 & 4.55e-02 & 9.54e-01 & 1.02e-06 & 1.44e-05 & 5.19e-08 \\
145 & C$_{\rm dep}$ & 3.15e+00 & 1.24e+02 & 1.07e+02 & 1.00e+02 & 6.98e+00 & 4.95e+00 & 6.37e+02 & 5.92e-01 & 1.26e-01 & 8.74e-01 & 3.67e-07 & 1.58e-05 & 3.28e-08 \\
150 & C$_{\rm dep}$ & 3.16e+00 & 1.28e+02 & 1.19e+02 & 1.10e+02 & 6.97e+00 & 5.20e+00 & 1.50e+03 & 9.13e-01 & 5.29e-02 & 9.47e-01 & 6.43e-07 & 1.52e-05 & 3.53e-08 \\
\hline\hline
\end{tabular}
\caption{
\label{tab:WR_feh_m3_tab}
Same as \autoref{tab:WR_feh_m2_tab} for [Fe/H]=-3.0 ([$\alpha$/Fe]=0.4).
}
\end{table}
\end{landscape}

\begin{landscape}
\begin{table}
\centering
\begin{tabular}{ccccccccccccccc}
\hline\hline
$M_{\rm{initial}}$ & 
Final &
$t_{\rm halt}$ &
$M_{\rm final}$ &
$M_{\rm He,core}$ &
$M_{\rm CO,core}$ &
$\log L$ &
$\log T_{\rm eff}$ &
$v_{\rm crit,eq}$ &
$v/v_{\rm crit,eq}$ &
$X_{\rm H,surf}$ &
$X_{\rm He,surf}$ &
$X_{\rm C,surf}$ &
$X_{\rm N,surf}$ &
$X_{\rm O,surf}$ \\
(M$_\odot$) &
burning &
(Myr) &
(M$_\odot$) &
(M$_\odot$) &
(M$_\odot$) &
(L$_\odot$) &
(K) &
(km s$^{-1}$) \\
\hline
10 & O$_{\rm dep}$ & 2.60e+01 & 9.97e+00 & 3.07e+00 & 2.54e+00 & 4.75e+00 & 3.63e+00 & 6.57e+01 & 1.87e-02 & 7.16e-01 & 2.84e-01 & 1.08e-08 & 1.26e-06 & 4.02e-07 \\
15 & C$_{\rm dep}$ & 1.43e+01 & 1.50e+01 & 4.73e+00 & 3.30e+00 & 4.81e+00 & 3.70e+00 & 9.09e+01 & 1.63e-01 & 7.40e-01 & 2.60e-01 & 1.16e-08 & 1.14e-06 & 5.40e-07 \\
20 & C$_{\rm dep, \, marginal}$ & 1.02e+01 & 1.98e+01 & 5.99e+00 & 4.06e+00 & 5.22e+00 & 4.34e+00 & 2.81e+02 & 8.29e-01 & 7.28e-01 & 2.72e-01 & 8.23e-09 & 1.24e-06 & 4.36e-07 \\
25 & C$_{\rm dep, \, marginal}$ & 8.20e+00 & 2.46e+01 & 8.01e+00 & 5.30e+00 & 5.45e+00 & 4.26e+00 & 2.09e+02 & 8.26e-01 & 7.14e-01 & 2.86e-01 & 7.10e-09 & 1.27e-06 & 4.04e-07 \\
30 & C$_{\rm dep}$ & 7.35e+00 & 2.92e+01 & 1.04e+01 & 7.10e+00 & 5.67e+00 & 4.26e+00 & 2.06e+02 & 9.51e-01 & 6.60e-01 & 3.40e-01 & 6.28e-09 & 1.36e-06 & 3.02e-07 \\
35 & C$_{\rm dep}$ & 7.80e+00 & 3.32e+01 & 2.31e+01 & 1.91e+01 & 5.99e+00 & 4.16e+00 & 1.31e+02 & 1.97e-01 & 3.67e-01 & 6.33e-01 & 1.33e-08 & 1.94e-06 & 9.35e-08 \\
40 & C$_{\rm dep}$ & 6.98e+00 & 3.77e+01 & 2.81e+01 & 2.35e+01 & 6.12e+00 & 4.23e+00 & 1.46e+02 & 1.34e-01 & 3.54e-01 & 6.46e-01 & 4.34e-08 & 3.15e-06 & 9.27e-08 \\
45 & C$_{\rm dep, \, marginal}$ & 6.12e+00 & 4.28e+01 & 2.65e+01 & 2.20e+01 & 6.12e+00 & 4.40e+00 & 2.55e+02 & 4.82e-01 & 3.64e-01 & 6.36e-01 & 4.42e-08 & 3.56e-06 & 1.00e-07 \\
50 & C$_{\rm dep, \, marginal}$ & 5.34e+00 & 4.68e+01 & 2.72e+01 & 1.30e+01 & 6.07e+00 & 4.19e+00 & 1.62e+02 & 4.69e-01 & 4.72e-01 & 5.28e-01 & 7.55e-09 & 1.51e-06 & 1.28e-07 \\
55 & C$_{\rm dep}$ & 5.15e+00 & 4.94e+01 & 3.74e+01 & 3.31e+01 & 6.24e+00 & 4.48e+00 & 2.78e+02 & 2.85e-01 & 3.01e-01 & 6.99e-01 & 8.67e-09 & 1.59e-06 & 3.68e-08 \\
60 & C$_{\rm dep}$ & 4.91e+00 & 5.21e+01 & 4.34e+01 & 3.83e+01  & 6.32e+00 & 4.56e+00 & 3.00e+02 & 7.43e-01 & 2.62e-01 & 7.38e-01 & 8.75e-09 & 1.59e-06 & 3.21e-08 \\
65 & C$_{\rm dep}$ & 4.71e+00 & 5.66e+01 & 5.02e+01 & 4.41e+01 & 6.42e+00 & 4.97e+00 & 8.62e+02 & 1.30e+00 & 1.22e-01 & 8.78e-01 & 1.50e-08 & 2.39e-06 & 1.28e-08 \\
70 & C$_{\rm dep}$ & 4.46e+00 & 6.23e+01 & 5.17e+01 & 4.82e+01 & 6.39e+00 & 4.56e+00 & 2.87e+02 & 5.33e-01 & 2.25e-01 & 7.75e-01 & 9.12e-09 & 1.60e-06 & 1.85e-08 \\
75 & C$_{\rm dep}$ & 4.34e+00 & 4.35e+01 & 4.35e+01 & 3.74e+01 & 6.38e+00 & 5.45e+00 & 1.19e+03 & 3.62e-01 & 8.94e-25 & 1.18e-01 & 3.34e-01 & 2.40e-15 & 5.03e-01 \\
80 & C$_{\rm dep}$ & 4.16e+00 & 7.19e+01 & 6.11e+01 & 5.62e+01 & 6.51e+00 & 4.62e+00 & 3.06e+02 & 5.39e-01 & 2.03e-01 & 7.97e-01 & 9.58e-09 & 1.61e-06 & 1.51e-08 \\
85 & C$_{\rm dep}$ & 3.96e+00 & 7.56e+01 & 6.02e+01 & 5.40e+01 & 6.46e+00 & 4.56e+00 & 2.99e+02 & 4.65e-01 & 2.43e-01 & 7.57e-01 & 9.15e-09 & 1.60e-06 & 1.90e-08 \\
90 & C$_{\rm dep}$ & 3.88e+00 & 8.12e+01 & 6.54e+01 & 5.84e+01 & 6.51e+00 & 4.59e+00 & 3.03e+02 & 4.39e-01 & 2.35e-01 & 7.65e-01 & 9.26e-09 & 1.60e-06 & 1.95e-08 \\
95 & C$_{\rm dep}$ & 3.78e+00 & 8.59e+01 & 7.02e+01 & 5.84e+01 & 6.56e+00 & 4.61e+00 & 3.21e+02 & 4.83e-01 & 2.16e-01 & 7.84e-01 & 9.40e-09 & 1.61e-06 & 1.47e-08 \\
100 & C$_{\rm dep}$ & 3.74e+00 & 7.66e+01 & 7.66e+01 & 6.40e+01 & 6.69e+00 & 5.48e+00 & 1.30e+03 & 1.17e-01 & 2.95e-24 & 8.19e-02 & 2.18e-01 & 1.94e-09 & 6.14e-01 \\
105 & C$_{\rm dep}$ & 3.62e+00 & 9.47e+01 & 7.74e+01 & 7.01e+01 & 6.72e+00 & 4.91e+00 & 6.49e+02 & 6.06e-01 & 1.54e-01 & 8.46e-01 & 2.24e-08 & 1.59e-06 & 1.09e-08 \\
110 & He$_{\rm dep}$ & 3.54e+00 & 1.00e+02 & 8.21e+01 & 7.44e+01 & 6.72e+00 & 4.85e+00 & 6.00e+02 & 6.00e-01 & 1.61e-01 & 8.39e-01 & 2.36e-08 & 1.59e-06 & 1.10e-08 \\
115 & C$_{\rm dep, \, marginal}$ & 3.47e+00 & 1.05e+02 & 8.39e+01 & 3.15e+01 & 6.73e+00 & 4.75e+00 & 4.39e+02 & 4.18e-01 & 1.61e-01 & 8.39e-01 & 2.29e-08 & 1.59e-06 & 1.18e-08 \\
120 & C$_{\rm dep, \, marginal}$ & 3.42e+00 & 1.10e+02 & 8.51e+01 & 4.61e+01 & 7.16e+00 & 4.49e+00 & 1.71e+02 & 2.24e-01 & 1.48e-01 & 8.52e-01 & 3.91e-07 & 5.93e-06 & 1.23e-05 \\
125 & C$_{\rm dep, \, marginal}$ & 3.37e+00 & 1.16e+02 & 8.79e+01 & 7.69e+01 & 7.17e+00 & 4.50e+00 & 1.77e+02 & 1.89e-01 & 1.39e-01 & 8.61e-01 & 2.78e-08 & 1.59e-06 & 3.37e-09 \\
130 & C$_{\rm dep, \, marginal}$ & 3.28e+00 & 1.21e+02 & 9.19e+01 & 7.68e+01 & 1.20e+01 & 5.94e+00 & 7.96e+01 & 1.27e+00 & 2.58e-01 & 7.42e-01 & 1.02e-08 & 1.60e-06 & 1.95e-08 \\
135 & C$_{\rm dep, \, marginal}$ & 3.25e+00 & 1.27e+02 & 9.32e+01 & 9.22e+01 & 7.27e+00 & 4.45e+00 & 1.58e+02 & 1.43e-01 & 2.12e-01 & 7.88e-01 & 1.65e-08 & 1.60e-06 & 1.44e-08 \\
140 & C$_{\rm dep, \, marginal}$ & 3.22e+00 & 1.31e+02 & 1.00e+02 & 6.56e+01 & 7.57e+00 & 4.31e+00 & 8.94e+01 & 1.23e-01 & 1.36e-01 & 8.64e-01 & 6.30e-08 & 1.79e-06 & 1.07e-08 \\
145 & C$_{\rm dep, \, marginal}$ & 3.17e+00 & 1.34e+02 & 1.07e+02 & 8.16e+01 & 6.94e+00 & 4.95e+00 & 7.96e+02 & 7.63e-01 & 1.40e-01 & 8.60e-01 & 5.12e-08 & 1.54e-06 & 4.12e-09 \\
150 & C$_{\rm dep, \, marginal}$ & 3.14e+00 & 1.40e+02 & 1.06e+02 & 8.18e+01 & 7.03e+00 & 4.56e+00 & 2.40e+02 & 4.38e-01 & 1.35e-01 & 8.65e-01 & 5.14e-08 & 1.54e-06 & 4.14e-09 \\
\hline\hline
\end{tabular}
\caption{
\label{tab:WR_feh_m4_tab}
Same as \autoref{tab:WR_feh_m2_tab} for [Fe/H]=-4.0 ([$\alpha$/Fe]=0.4).
}
\end{table}
\end{landscape}

\subsection{Evolutionary paths}
\label{model_grid_evol_sec}

In this section, we begin by examining what classes of stars our model grids produce. For reference, we tabulate the stellar properties at the final time that we reach for each model in \autoref{tab:WR_feh_m2_tab}, \autoref{tab:WR_feh_m3_tab} and \autoref{tab:WR_feh_m4_tab} for [Fe/H]$=-2.0$, $-3.0$, $-4.0$, respectively, and for our fiducial rotation rate $v/v_{\rm crit}=0.4$. A first, basic question is what classes of stars our models produce. We summarise the classes of interest to us, and how we define them in the context of our model grid, in \autoref{tab:def_tab_WR}. The categories we consider are:

\begin{itemize}
\item \textbf{O stars.} As discussed in \citet{roy2019}, O stars are defined as those which are core H burning, $X_{\rm H,c} > 10^{-4}$, and their surfaces are not yet contaminated by the nucleosynthetic elements, and therefore the surface He abundance is $X_{\rm He,surf} < 0.4$.      
\item \textbf{WNE and WNL stars.} As discussed in \citet{roy2019}, rapidly-rotating massive stars are likely to evolve into Wolf-Rayet stars of the WN subclass, even in the absence of strong winds, due to rotationally-driven dredge-up. Following that paper, we classify these stars based on their surface N, C, and He abundances; specifically, we take WNLs to be stars with $0.4 < X_{\rm He,surf} < 0.9$ and $X_{\rm C,surf}/X_{\rm N,surf} < 10$, while WNEs have $X_{\rm He,surf} > 0.9$ and $X_{\rm C,surf}/X_{\rm N,surf} < 10$. These criteria are approximate -- Wolf-Rayet subtypes are ultimately assigned based on spectroscopic characteristics -- but \citet{roy2019} carry out stellar atmosphere modelling for some of their model stars, and show that these surface abundance criteria are generally consistent with spectroscopic classifications.
\item \textbf{WC and WO stars.} WC stars are generally though to be core He burning stars that have significant amounts of C on their surfaces as a result of envelope loss or rotational mixing; late in core He burning, these may evolve into WO stars as significant O appears on the surface as well. Their origins have been studied extensively by \citet{crowther1998}, \citet{nugis2000}, and \citet{crowther2007}. We assign stars to these classes following \citet{crowther2007}; specifically, we define WC stars as those with surface abundances $0.1 < X_{\rm C,surf} < 0.6$, $X_{\rm O,surf} < 0.1$ and central H and He abundances $X_{\rm H,c} < 10^{-4}$ and $X_{\rm He,c} > 0.1$. Similarly, we define WO stars to be those with surface abundances obeying $X_{C,\rm surf} > 0.1$, $X_{\rm O,surf} > 0.1$, and $(X_{\rm C,surf} + X_{\rm O,surf})/X_{\rm He,surf} > 1$, and central abundances $X_{\rm H,c} < 10^{-4}$ and $X_{\rm He,c} < 0.1$. Again, we caution that these are approximations, since neither we nor previous authors have carried out the stellar atmosphere modelling that would be required to match these core and surface characteristics to observable spectral features.
\end{itemize}

\begin{figure}
\centerline{
\includegraphics[width=0.51\textwidth]{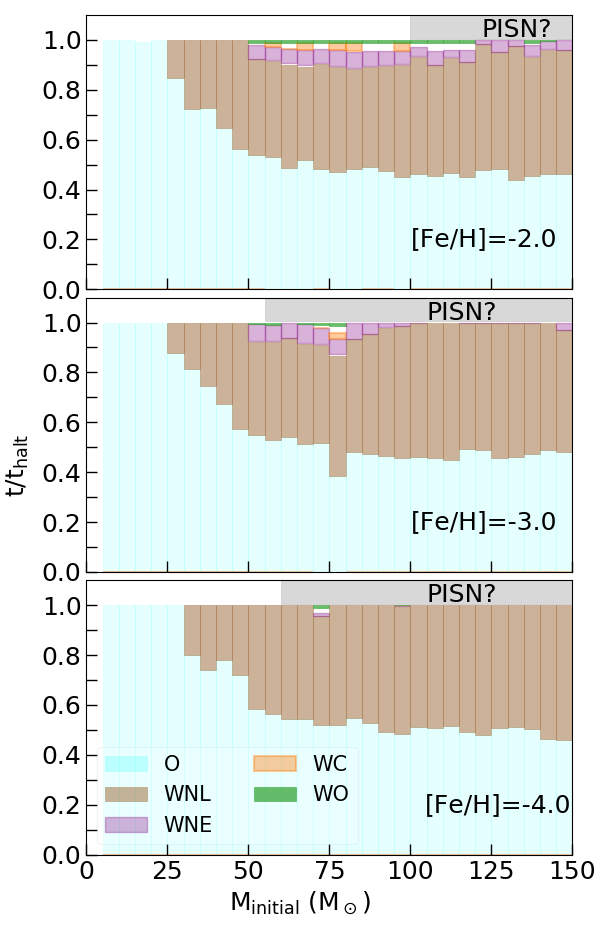}
}

\caption{
Classifications of stars of a specified initial mass (horizontal axis) as a function of time, normalised to the halting time $t_{\rm halt}$ at which we end the simulation of that model (see \autoref{tab:WR_feh_m2_tab} - \autoref{tab:WR_feh_m4_tab}); colour indicates the classification, with stars that are identified as O stars, WNL, WNE, WC, WO. The models shown are for our fiducial rotation rate $v/v_{\rm crit} = 0.4$, and the three panels show three different metallicities, as indicated. The grey band at the top, labelled ``PISN?'', indicates stars that might produce pair instability SNe.
}
\label{fig:WRsub04_mod}
\end{figure}

We run our models until the end of core O depletion for the least masive stars (10 M$_\odot$) in our model grids to ensure that they successfully go through the O burning in a sufficiently non-degenerate core, which will lead to hydrostatic core Si burning and eventually to SN. Moreover, this implies that all our more massive stars ($>10 \, \mathrm{M}_\odot$) will also successfully go through the non-degenerate core O burning followed by Si burning, and then to SN or direct collapse to Black Holes. Therefore, we conclude that all stars in our model grids are ``true" massive stars, and for more massive stars ($>10 \, \mathrm{M}_\odot$), we can safely halt our simulations at the end of core C depletion without proceeding further into the core O, Si - burning phases. We summarise the application of these classifications to our model grid for our fiducial rotation rate, $v/v_{\rm crit} = 0.4$, in \autoref{fig:WRsub04_mod}. In this figure, the horizontal axis shows the initial mass, and the vertical axis shows time relative to the halting time shown in \autoref{tab:WR_feh_m2_tab}, \autoref{tab:WR_feh_m3_tab} and \autoref{tab:WR_feh_m4_tab}, the time at which the simulation ends; since we run through the end of C burning, this should be nearly identical to the full stellar lifetime. Colours then indicate the classification for a star of the indicated mass and evolutionary time in our grid. We can make a few immediate observations. First, for the mass range we have explored, all models remain sufficiently non-degenerate to proceed through Si burning, and thence either to SN or to direct collapse to black hole. Second, WR phases begin to appear at initial masses $\gtrsim 30$ $M_\odot$. Stars in this mass range spend a significant amount of time as WNL stars, small amounts of time as WNEs, and only tiny periods as WCs or WOs. The relative dearth of WC and WO subclasses is not surprising given the very metal-poor populations that we are modelling, which experience comparatively little mass loss. By contrast with these types, the WN subtypes can be produced by rotational dredge-up, and thus can be produced even at low metallicity \citep{roy2019}. Stars below $\sim 30$ $M_\odot$ show characteristics of massive main-sequence O stars.

Although the figure shows only our fiducial rotation rate case, we note that the results for more rapid rotation ($v/v_{\rm crit} = 0.6$) are qualitatively very similar. For our lowest rotation rate case ($v/v_{\rm crit} = 0.2$), we produce far fewer WN stars due to the lack of rotational mixing -- see \citet{roy2019} for a detailed discussion.

It is also of interest to ask how our models will end their lives. Since we have not run all the way to a pre-SN state, we can only diagnose this indirectly, based on the CO core mass and comparison to other model grids that have been run further. \citet{limongi2018} find that stars with CO core masses $\gtrsim 35$ M$_\odot$ may end their lives at pair instability SNe, and we label the models that satisfy this condition as PISN in \autoref{fig:WRsub04_mod}.\footnote{\citeauthor{limongi2018}'s model grid goes to a minimum metallicity $\mathrm{[Fe/H]} = -3$, but we assume that this result also applies to our $\mathrm{[Fe/H]} = -4$ models, since both CO and He core masses depend very weakly on metallicity.} However, we caution that this identification is extremely uncertain, and that these stars may simply collapse directly to black holes instead, as suggested by \citet{sukhbold16}.

\begin{table*}
\centerline{
\begin{tabular}{ccc}
\hline\hline
Classification & Surface state &  Central state \\
\hline
O stars & $X_{\rm He} < 0.4$ &  $X_{\rm{H}}>10^{-4}$ \\
WNL & $X_{\rm He}= 0.4-0.9 \qquad X_{\rm C}/X_{\rm N} <10$  & - \\
WNE & $X_{\rm He} > 0.9 \qquad X_{\rm C}/X_{\rm N} <10$  & - \\
WC & $0.1<X_{\rm C}<0.6 \qquad X_{\rm O}<0.1 \qquad X_{\rm C}/X_{\rm N} >10$ 
& $X_{\rm{H}}<10^{-4} \qquad X_{\rm{He}}>0.1$\\
WO & $X_{\rm C}>0.1 \qquad X_{\rm O}>0.1 \qquad (X_{\rm C}+X_{\rm O})/X_{\rm He} > 1$ & $X_{\rm{H}}<10^{-4} \qquad X_{\rm{He}}<0.1$\\
\hline\hline
\end{tabular}
}
\caption{
\label{tab:def_tab_WR}
Criteria we use to classify stars as ``ordinary" O, WNL, WNE, WC, and WO. Surface state refers to conditions at the stellar surface, and core state to conditions at the centre of the star; $X_{\rm Q}$ indicates the mass fraction of element Q. See main text for references and explanations.
}
\end{table*}

\subsection{Time evolution of yields, and the origin of N}
\label{time_yield_sec}

\begin{figure*}
\centerline{
\includegraphics[width=1.0\textwidth]{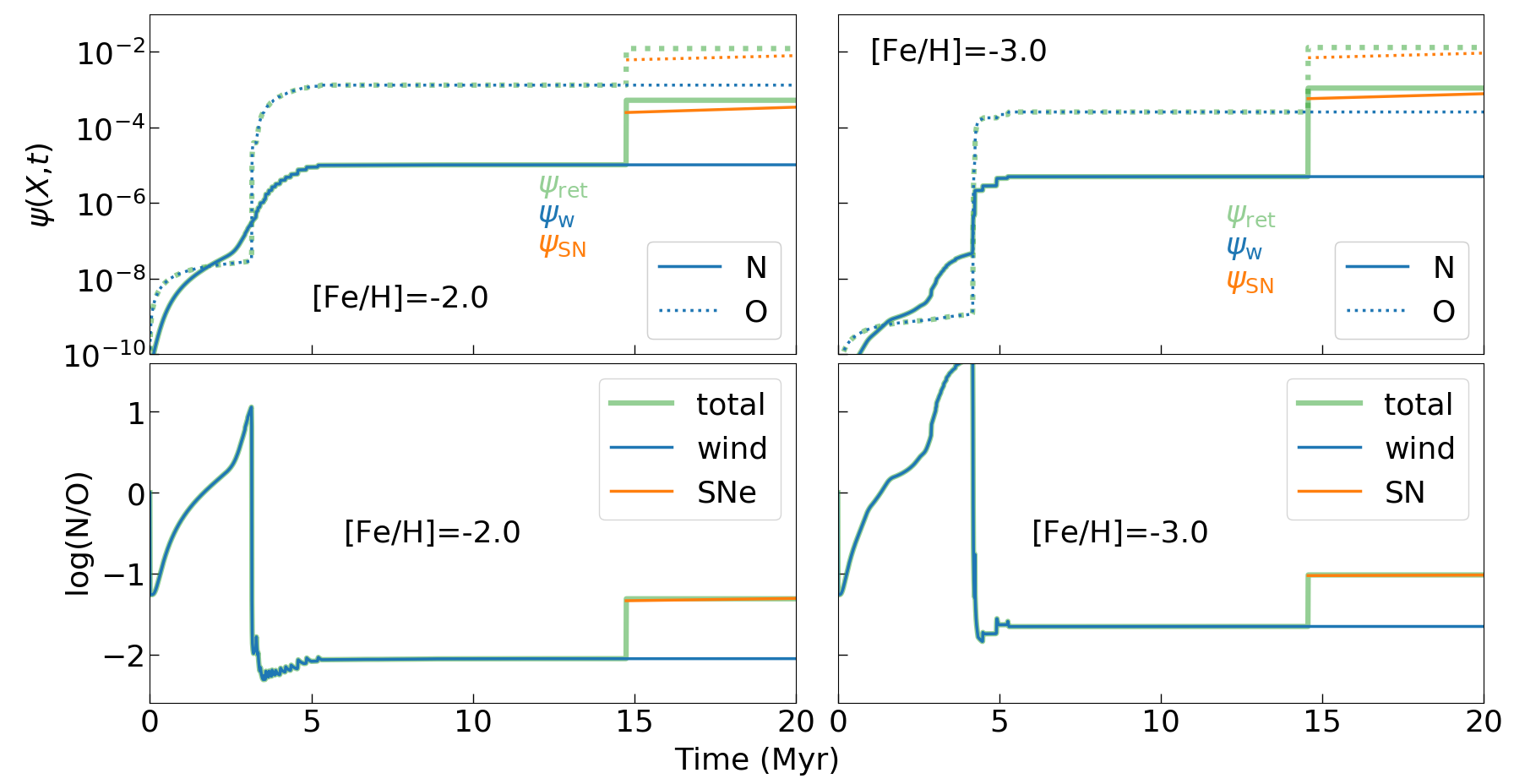}
}

\caption{{\it Top panels}: The time evolution of the return fraction $\psi_{\rm ret}(X, t)$  for N (solid lines) and O (dotted lines) coming from winds  (blue lines) and from SNe ejecta (orange). {\it Bottom panels}: Time evolution of N/O coming from pre-SN winds (blue solid line) and SN ejecta. {\it Left panels}: [Fe/H]=-2.0 ([$\alpha$/Fe]=0.4). {\it Right panels}: [Fe/H]=-3.0 ([$\alpha$/Fe]=0.4). All results are for stars with $v/v_{\rm{crit}}=0.4$.
}
\label{fig:IMF_NO}
\end{figure*}

\begin{figure}
\centerline{
\includegraphics[width=0.5\textwidth]{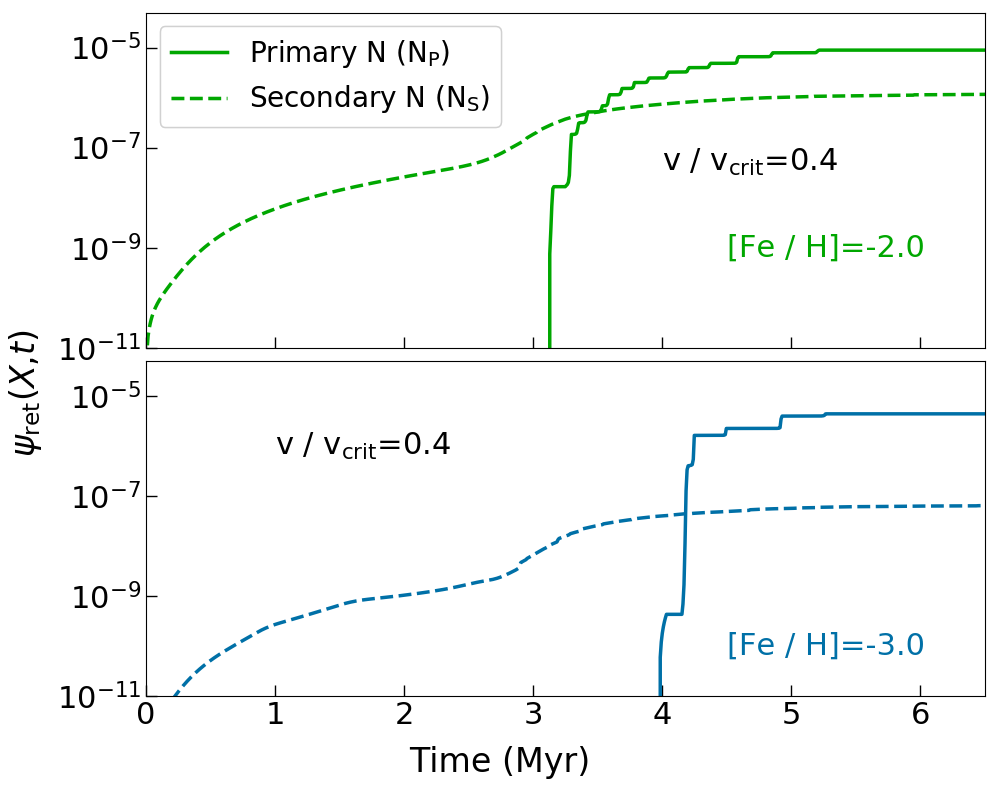}
}

\caption{Time evolution of cumulative nitrogen mass return from stellar winds $\psi_{\rm w}(X,t)$ ($=\psi_{\rm ret}(X,t)$ for $\psi_{\rm SN}(X,t)=0.0$) for primary (N$_{\rm P}$) and secondary N (N$_S$) for two metallicities; the two cases shown are the same as in \autoref{fig:IMF_NO}: $[\mathrm{Fe}/\mathrm{H}]=-2.0$ (green, top panel), $-3.0$ (blue, bottom panel).  The solid and dashed lines show primary nitrogen, N$_{\rm P}$, and secondary nitrogen, N$_S$, respectively.  
}
\label{fig:NP_NS}
\end{figure}

\begin{figure}
\centerline{
\includegraphics[width=0.5\textwidth]{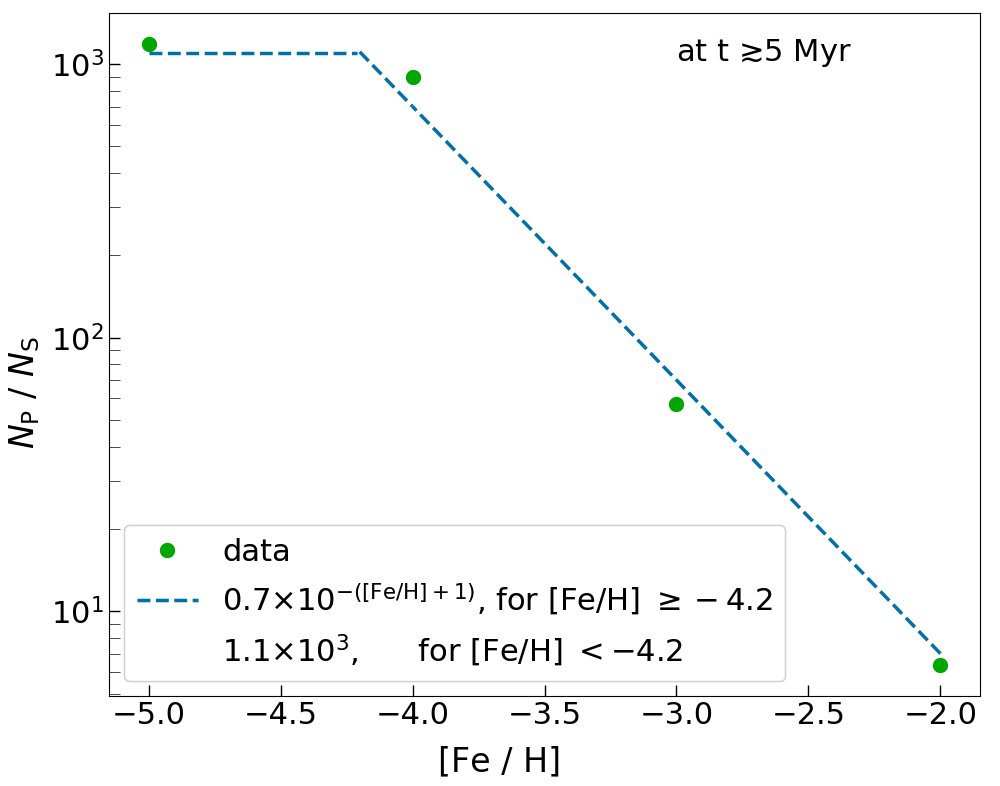}
}

\caption{N$_{\rm P}$/N$_S$ versus [Fe/H] for [Fe/H] $\leq -2.0$ at a time $\gtrsim 5$ Myr. The green circles represent the simulation data and the blue dashed line represents a fit: $0.7 \times 10^{-([\mathrm{Fe/H}]+1)}$ for [Fe/H] $\geq -4.0$ and 1.1$\times 10^3$ otherwise. 
}
\label{fig:NPNS_feh}
\end{figure}

Having discussed the model grids in \autoref{model_grid_evol_sec}, in this section we discuss the time evolution of yields, and to do that, we first examine two representative metallicity cases, $[\mathrm{Fe/H}]=-2$ and $-3$, for $v/v_{\rm crit} = 0.4$, in order to establish general patterns of chemical enrichment and the physics that drive them. We show the time evolution of the mass return $\psi_{X,\rm ret}(t)$ for N and O in \autoref{fig:IMF_NO}. In this figure, the top panels show the mass return due to winds and SNe, and the bottom panels show the N to O ratio in the ejecta produced by both sources. The general pattern we find is that the N/O ratio from wind is initially high due to ejection of CNO-processed material that has been dredged to the surface by rotation, but this ratio drops rapidly to $\log (\mathrm{N/O})\sim -2$ to $-1.5$ due to the ejection of O-rich material immediately at and after core H depletion, during the WR evolutionary phase. All of this occurs prior to the first SN. WR winds constitute the majority of the mass return prior to the onset of SNe. Once SNe begin, they completely dominate the mass budget, and, after a short transient, also produce ejecta with abundance ratio $\log (\mathrm{N/O})\sim -1.25$ to $-1.0$. This ratio appears to be characteristic of all mass return from massive stars. 

We also examine the nucleosynthetic origin of N from massive stars in \autoref{fig:NP_NS}, which separates the cumulative nitrogen mass return from winds $\psi_{w}(X,t)$ into primary (N$_P$) and secondary (N$_S$) for the same metallicities and rotation rates as in \autoref{fig:IMF_NO}. In order to explain the procedure by which we construct this figure, we must review the processes that produce N. When the core is CNO burning, N is secondary because it is produced by the conversion of C which was already present in the star at birth. The secondary N then gets dredged-up to the stellar surfaces by rotationally induced mixing and subsequently gets ejected to the ISM by the MS wind mass-loss. On the other hand, when the core is He burning, the triple-$\alpha$ process produces C that can also be transported upward by rotationally-induced mixing and then be converted to N by the CNO-process during H-shell burning. This channel of N production is primary because the C that is being converted to N is self-produced rather than inherited from the ISM.
Consistent with this discussion, we determine the origin of the N ejected into stellar winds in our evolutionary models as follows: at each time step, we examine both the state of nuclear burning in the core (i.e., is the core H- or He-burning) and the total mole fraction of C ($^{12} \mathrm{C}$), N ($^{14} \mathrm{N}$), and O ($^{16} \mathrm{O}$), $Y_{\rm C} + Y_{\rm N} + Y_{\rm O}$, in the star. If the core is H-burning, then we are during the phase when the star has not yet produced any C of its own, and we classify any N lost as secondary (${\rm N}_{\rm S}$). If the core is He-burning, but the total mole fraction of C, N, and O in the star is less than or equal to the initial mole fraction (i.e., $Y_{\rm C} + Y_{\rm N} + Y_{\rm O} \approx Y_{\rm C,init} + Y_{\rm N,init} + Y_{\rm O,init}$), then the star still has not produced significant primary C, and thus any N must again be secondary. We classify N as primary only if the star is He burning \textit{and} the total mole fraction of C, N, and O exceeds the initial fraction (i.e., $Y_{\rm C} + Y_{\rm N} + Y_{\rm O} > Y_{\rm C,init} + Y_{\rm N,init} + Y_{\rm O,init}$); this second condition ensures that the star has begun to self-enrich with primary C, which is then being converted to primary N.

We see from \autoref{fig:NP_NS} that secondary N production dominates for the first $\approx 3-4$ Myr after a stellar population forms, after which primary production begins; at late times the ratio of primary to secondary production depends on the initial metallicity, with secondary production dominating above $[\mathrm{Fe}/\mathrm{H}] \gtrsim -1$, and primary at lower metallicities. We find that, before $\sim 3.3 - 4.0$ Myr for $[\mathrm{Fe}/\mathrm{H}]=-2$ and $-3$ respectively, N is purely secondary, with primary becoming dominant after $\sim 4- 5$ Myr.

We plot the ratio of primary to secondary N, ${\rm N}_P/{\rm N}_S$, at late times ($\gtrsim 5$ Myr) as a function of iron metallicity $[{\rm Fe}/{\rm H}]$ in
\autoref{fig:NPNS_feh} for the same rotation rate as shown in \autoref{fig:IMF_NO}. Note that we have added a point at $[{\rm Fe}/{\rm H}]=-5$ that is not part of our full grid, but is computed in the same way, in order to extend the correlation.
The green circles show the data at four metallicities and the blue dashed line shows the following fit:
\begin{equation}
\mathrm{N}_{\mathrm{P}}/\mathrm{N}_{\mathrm{S}} =
\left\{
\begin{array}{ll}
0.7 \times 10^{-([\mathrm{Fe/H}]+1)}, & [\mathrm{Fe}/\mathrm{H}] \geq -4.0 \\
1.1 \times 10^3, & [\mathrm{Fe}/\mathrm{H}] < -4.0
\end{array}
\right..
\label{eq:NPNS_feh_fit}
\end{equation} 
We find that as the metallicity decreases, N$_P$/N$_S$ increases following the first relation in \autoref{eq:NPNS_feh_fit}, and at very low metallicities, [Fe/H] $\lesssim -4.0$, N$_P$/N$_S$ almost saturates at $\sim 1000$ as shown in \autoref{fig:NPNS_feh}. Thus, over the metallicity range we consider, N is completely dominated by primary production.

\subsection{Ejecta velocity and retention}
\label{ssec:velocity}

\begin{figure}
\centerline{
\includegraphics[width=0.5\textwidth]{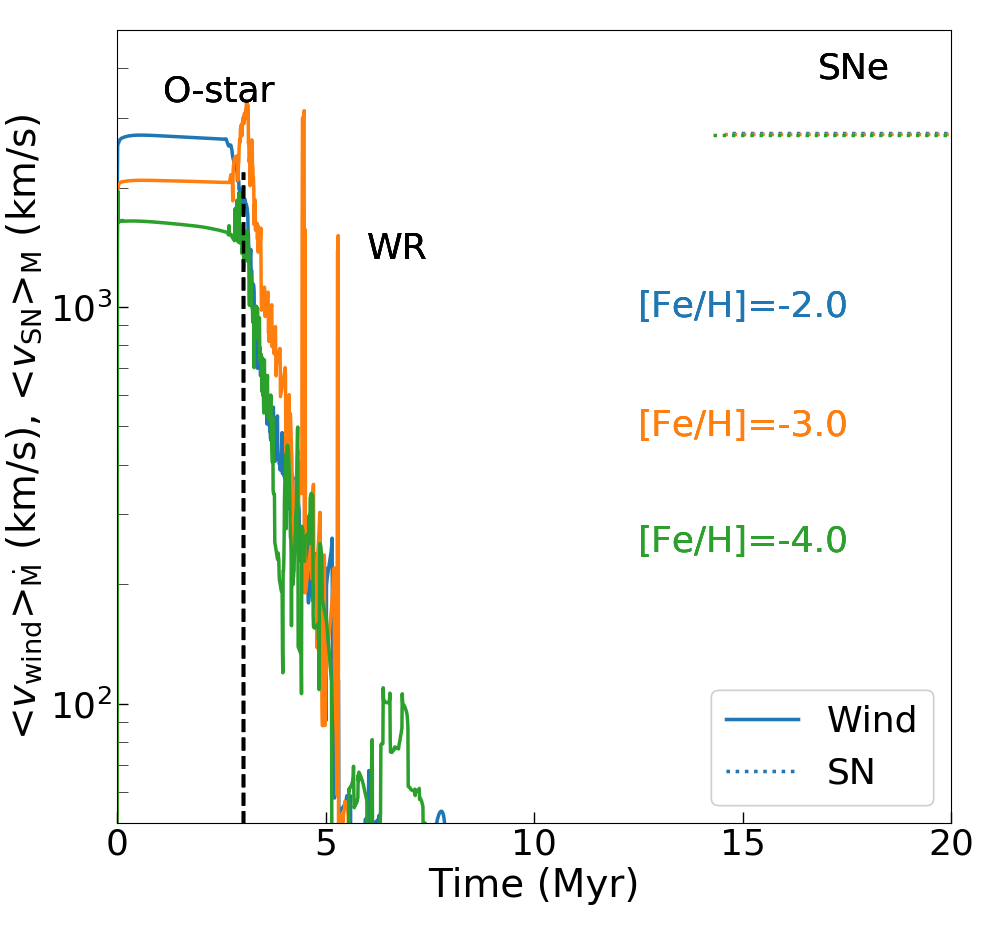}
}

\caption{Time evolution of mass-loss weighted mean wind velocity $\langle v_{\rm wind}\rangle _{\dot{M}}$ for three metallicities, $[\mathrm{Fe}/\mathrm{H}]=-2.0$, $-3.0$, $-4.0$ and for $v/v_{\rm crit}=0.4$ (solid lines), and mass-weighed mean SN ejecta velocity $\langle v_{\rm SN}\rangle _{M}$ (dotted lines) for $[\mathrm{Fe}/\mathrm{H}]=-2.0$ and $-3.0$). The black dashed vertical line marks the approximate time at which the most massive stars enter the WR phase. 
}
\label{fig:vwind}
\end{figure}

\begin{figure*}
\centerline{
\includegraphics[width=1.0\textwidth]{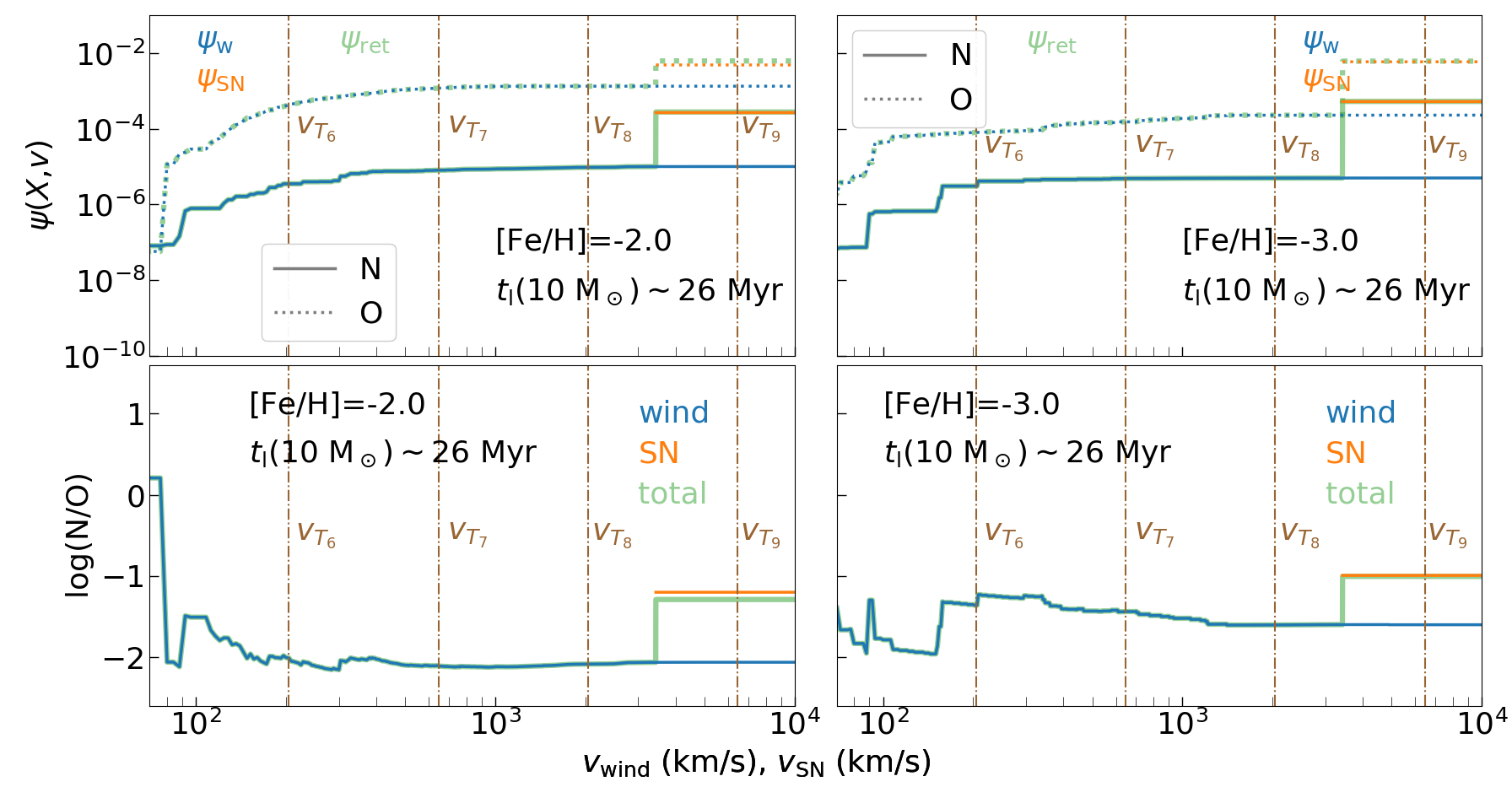}
}

\caption{ {\it Top panels}: Return fraction $\psi_{\rm ret}(X, v)$ of material ejected with velocity $\leq v$, as defined by \autoref{eq:psi_v}, after the last SN explodes at $t_{\rm l}(10 \, \mathrm{M}_\odot) \sim 26$ Myr for N (solid lines) and O (dotted lines). We separate this into material coming from winds, $\psi_{\rm w}$ (blue lines), from SNe ejecta, $\psi_{\rm SN}$ (orange) and total, $\psi_{\rm ret}$ (green). {\it Bottom panels}: N/O ratio for material ejected at velocities $\leq v$, again separated into material coming from pre-SN winds (blue lines), SN ejecta (orange lines) and total (green lines). The cases shown in this figure are the same as those shown in \autoref{fig:IMF_NO}. The four vertical brown dashed-dotted lines represent velocities at temperatures $T=10^6$, $10^7$, $10^8$, $10^9$ K (denoted by $v_{T_x}$, where $T_x = 10^x$) respectively.
}
\label{fig:NO_v}
\end{figure*}

\begin{figure*}
\centerline{
\includegraphics[width=1.0\textwidth]{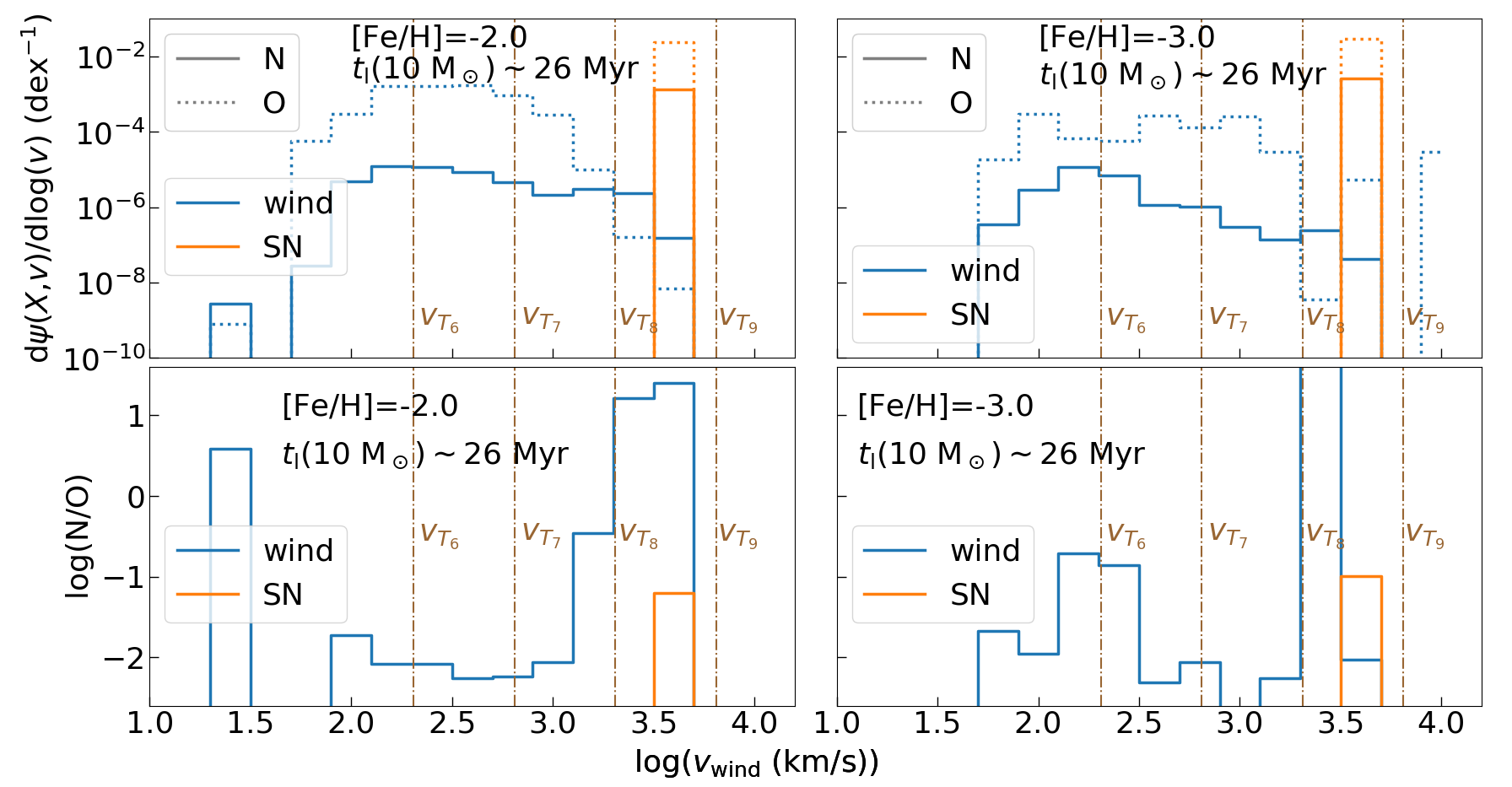}
}

\caption{ {\it Top panels}: Differential mass return ${\mathrm{d}\psi (X,v)}/{\mathrm{d}\log(v)}$ as a function of velocity integrated over bins 0.2 dex wide, as defined by \autoref{eq:diffPsi_dvbin}. Lines and colours are identical to those used in \autoref{fig:NO_v} {\it Bottom panels}: N/O ratio for material ejected in each of the velocity bins shown in the top panel.
The four vertical brown dashed-dotted lines represent velocities corresponding to particular temperatures, as in \autoref{fig:NO_v}.
}
\label{fig:dNO_dv}
\end{figure*}

We show the characteristic mass-loss weighted mean wind and mass-weighted mean SN ejecta velocity in \autoref{fig:vwind}. We see that winds (solid lines) have speeds of 1,000 - 2,000 km s$^{-1}$ during the O star phase and an order of magnitude lower during the WR phase. In contrast, typical ejecta velocities for SNe are typically $\sim 1.5 - 3.0$ times larger than the O star wind speed. This translates into an enormous difference in how easy it will be to retain the WR ejecta in a galaxy. The ejecta velocity is related to the temperature of the material once it shocks and halts by
\begin{equation}
T \approx \frac{m v^2}{3 N k_B} \,(\equiv \mu m_{\rm H} v^2/3k_B) = 2.4 \times 10^{5} \left(\frac{v}{100\,\mathrm{km\,s}^{-1}}\right)^2\mbox{ K} \, ,
\end{equation}
where $m$ is the ejecta mass, $N$ is the total number of particles amounting to mass $m$, and $\mu$ is the mean particle mass in H masses. We assume $\mu \sim 0.61$ for fully ionized gas.  
Thus WR wind ejecta will heat to just the peak of the cooling curve ($\sim 10^5$ K) and will therefore cool immediately. O-star wind-ejecta will be hotter by $\sim 2$ orders of magnitude ($\sim 10^7$ K), which is just past the peak of the cooling curve, and after undergoing a small amount of adiabatic expansion, much less than a factor of $\sim 2-3$ in radius, will undergo efficient radiative cooling. Supernova ejecta will be another order of magnitude hotter than that. Thus WR winds will never escape a galaxy, and it is at least plausible that O-wind ejecta (though these are negligible by mass compared to WR winds) will also be retained, since they will not be able to expand into a galactic halo before adiabatic expansion followed by radiative cooling saps their energy. By contrast, supernova ejecta will readily escape small galaxies. This hypothesis is strongly supported by simulations of SN in dwarf galaxies \citep[e.g.,][]{mac-low99a, emerick2018, emerick2019}, which show that dwarf galaxies retain almost none of their SN ejecta. By contrast, \citet{lochhaas2017} show that wind material can cool and form stars rapidly, in some cases even before the onset of the first SNe.  Moreover, wind ejecta encounter a largely unperturbed ISM, whereas SN ejecta pass through a more dilute ISM created by pre-supernova stellar feedbacks, further increasing the odds that SN ejecta will escape while wind ejecta will be retained.

In order to understand how differential retention of material ejecta at different velocities is likely to affect abundance patterns, we show the velocity-dependent return of N and O, $\psi_{\rm ret}({\rm N},v)$ and $\psi_{\rm ret}({\rm O},v)$, and the velocity-dependent abundance ratio ${\rm N}/{\rm O}(v)$, in \autoref{fig:NO_v}. We show the corresponding differential yield with respect to velocity, ${\rm d}\psi/{\rm d}\log v$, for these elements in \autoref{fig:dNO_dv}. These figures show the same cases as \autoref{fig:IMF_NO}, but here rather than showing how return varies over time, we now show how it varies with respect to ejecta velocity. We find that O, as a primary element, is produced and ejected via winds during the WR phase, and is therefore returned with comparatively low velocities of $\sim$ a few 100 km/s. The velocity-dependent return for O from winds therefore reaches its asymptotic value at a few 100 km/s. On the other hand, N, which has contributions from both primary and secondary processes, has a somewhat broader velocity distribution, with contributions at velocities of a few hundred km s$^{-1}$ similar to O, corresponding to primary N ejected during the WR phase when the core is He burning, but also some production at velocities $v \sim 1000$ km s$^{-1}$, corresponding to secondary N produced and ejected when the star is on the main sequence. As a result of these two contributions, the cumulative yield for N reaches its asymptotic value at a slightly larger velocity of $\sim 1000$ km/s. Nonetheless, we find that $\log({\rm N}/{\rm O})$ plateaus at $\sim -1.5$ to $-2.0$ for any velocity from $\approx 300-1,000$ km/s. In contrast, the return from SNe begins to appear above $\sim 3500$ km/s, and produce $\log({\rm N}/{\rm O}) \sim -1.0$. To help translate these into temperatures, in \autoref{fig:NO_v} we have added vertical lines indicating the velocities for which the post-shock temperature reaches the indicated values. We see that galaxies that retain material heated to $\sim 10^6 - 5 \times 10^7$ K will retain almost all of the N and O produced by winds, and will therefore have $\log({\rm N}/{\rm O})$ in the range $\sim -2$ to $\sim -1.5$, with very little dependence on the temperature. Galaxies that can retain material shock-heated to $\sim 5 \times 10^8$ K, by contrast, will hold onto their SN ejecta, and will therefore retain material with $\log({\rm N}/{\rm O})\sim -1$.

\begin{figure*}
\centerline{
\includegraphics[width=1.0\textwidth]{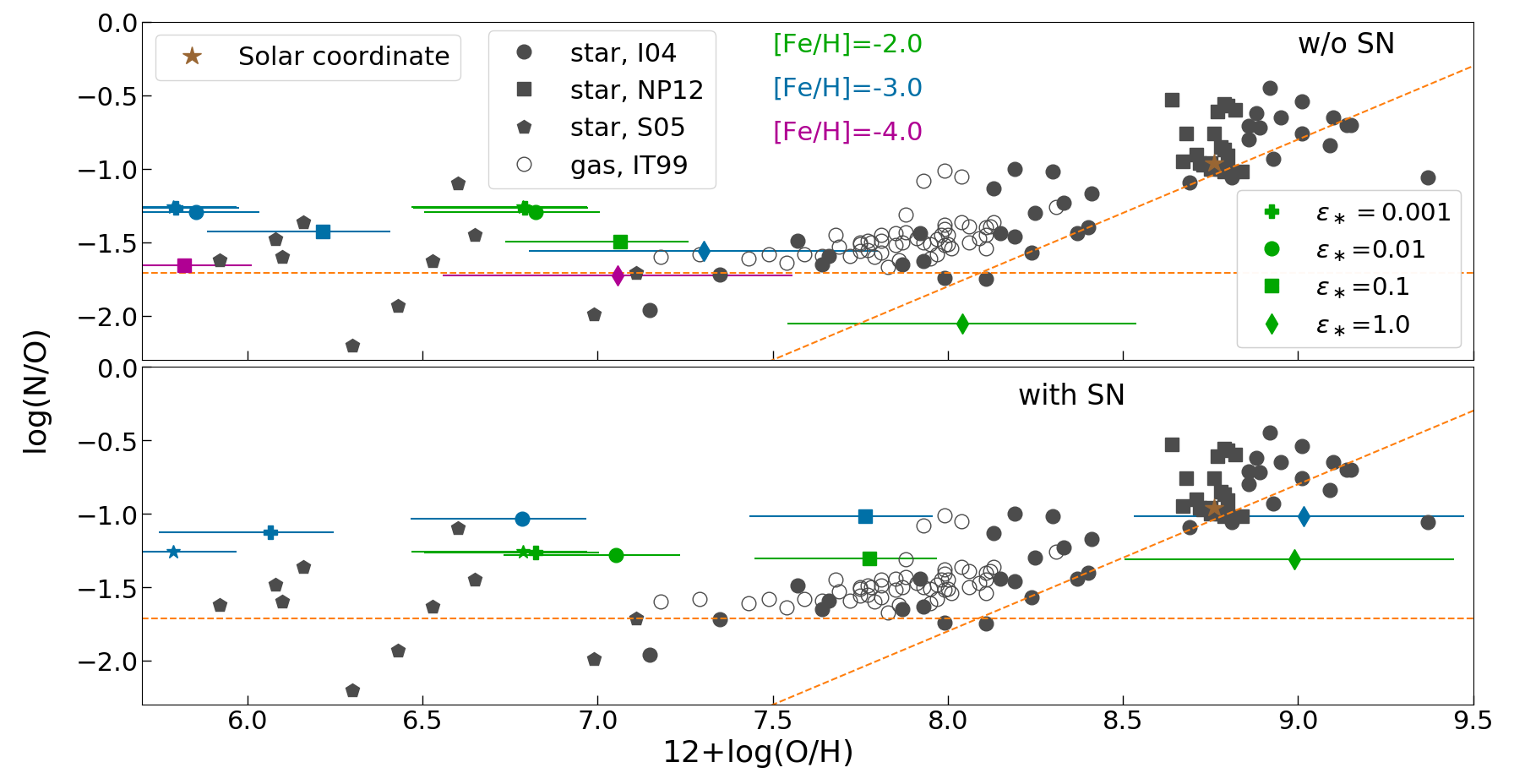}
}
\caption{$\log$(N/O) vs.~$\log$(O/H) as predicted by our massive star yield calculations, compared to observations. The two panels represent the two  different scenarios discussed in the main text: galaxies retain wind ejecta but lose all SN ejecta (top panel), galaxies retain both wind and SN ejecta and all stars $\leq 15$ M$_\odot$ explode (bottom panel). Black points are the same in  both panels, and show observations of N/O vs.~O/H in nearby \ion{H}{ii} regions (open circles; \citealt{izotov1999} (IT99)) and stars as denoted by filled markers (black):  \citealt{israelian2004} (I04, circles),  \citealt{spite2005} (S05, pentagons),  \citealt{nieva2012} (NP12, squares) ; orange dashed lines show the empirical primary and secondary production lines proposed by \citet{Dopita16}. The brown star shows the N/O and O/H ratios of the Sun. All other coloured points represent theoretical predictions of N/O vs.~O/H produced by nucleosynthesis by stellar populations with different initial iron metallicity $[\mathrm{Fe}/\mathrm{H}]$ and star formation efficiency $\epsilon_*$; green, blue, and purple show $[\mathrm{Fe}/\mathrm{H}] = -2$, $-3$, and $-4$, respectively, while different plot symbols show the star formation efficiencies $\epsilon_* = 0.001, 0.01, 0.1$, and $1$; for reference we also mark the point $\epsilon_* = 0$, in which case there is no star formation, and the point we are marking is simply the assumed initial state plus dilution by further accretion of primordial gas. For each point, the horizontal error bar shown corresponds to a galaxy that, subsequent to star formation, accretes an additional amount of primordial hydrogen and helium in an amount characterised by $\log \epsilon_{\rm p} = -0.5 - 0.5$. Note that, for the SN yields, we omit the case $[\mathrm{Fe}/\mathrm{H}]=-4$ because our SN yield tables only go down to $[\mathrm{Fe}/\mathrm{H}]=-3$.
}
\label{fig:NO_OH_key_fig}
\end{figure*}

\subsection{Distributions of N/O and O/H from massive star nucleosynthesis}

\label{yields_comp_sec}

Having discussed the general pattern of chemical enrichment in \autoref{time_yield_sec},
we now consider what distribution of N/O versus O/H we should observe if massive stars are the source of both N and O. To convert our calculations of N and O return as a function of initial [Fe/H] and rotation rate into a prediction for N/O and O/H, we consider a trivial chemical evolution model as described in \autoref{abund_sec}: a dwarf galaxy consisting of an initial gas reservoir converts some fraction of its gas into stars, which then return gas to the ISM, following which the galaxy also accretes some additional primordial material consisting of pure H and He (in the usual cosmic ratio). In this scenario, the ratios of any two elements can be computed by using \autoref{eq:Xi_Xj_pri}. The ingredients required for this calculation are the initial abundances in the reservoir $f_0$, the fraction $\epsilon_*$ of the reservoir mass to converted to stars, the mass return function $\psi_{\rm ret}$, and the  ratio of primordial mass accreted to reservoir mass $\epsilon_{\rm p}$. We can vary each of these parameters, and explore how they affect the resulting abundance ratios.

For the initial abundances, we consider galaxies with total iron metallicity $[\mathrm{Fe}/\mathrm{H}] = -4$, $-3$, and $-2$, scaling the abundances of all other elements relative to iron using the same empirical scaling we use for stars (\autoref{model_sec}). For star formation efficiency, we consider four possible values: $\log \epsilon_* = -3$, $-2$, $-1$, and 0 (i.e., 0.1\%, 1\%, 10\%, and 100\% of the initial reservoir is converted to stars). For additional accretion of primordial gas, we consider $\log\epsilon_{\rm p} = -0.5$ to $0.5$ (i.e., the reservoir subsequently accretes a factor of $\approx 0.3 - 3$ times its initial mass at the time of the starburst). Our motivation for choosing the relatively large value of $\epsilon_{\rm p} = 3$ is that observed dwarf galaxies (excluding those that have lost their gas due to environmental effects) are always gas-rich (e.g., \citealt{saintonge11a}).

The final parameter to specify is $\psi_{\rm ret}$, the mass return fraction from stellar nucleosynthesis. This will be dictated not just by nucleosynthesis itself, but by which nucleosynthetic products are retained in the galaxy rather than being lost. Given the large difference in ejecta velocity between winds and SNe, we consider two possibilities as far as what ejecta are retained by the galaxy. Our first scenario is to assume that, due to the comparatively low temperature of shocked winds as compared to SN ejecta, all wind material produced prior to the first SNe is retained in the galaxy, but that all SN ejecta or winds that injected after SNe begin are lost. Mathematically, in this scenario we take $M_{\rm SN}(m) = 0$ for all $m$, we set $\dot{M}_w(X,t)$ to the value returned by our stellar evolution grid for $t \leq t_\ell(15\,{\rm M}_\odot)$, where $t_\ell(15\,{\rm M}_\odot)$ is the lifetime of the first star to explode, and we set $\dot{M}_w(X,t) = 0$ for $t>t_\ell(15\,{\rm M}_\odot)$.\footnote{However, note that the results are nearly identical if we simply set $\dot{M}_w(X,t)$ to the value returned by our grid at all times, since the N/O ratio of the ejecta is nearly constant after $\approx 4$ Myr as shown in \autoref{fig:IMF_NO}.} Our second scenario is that both wind and SN ejecta are retained in the galaxy. For the purposes of this plot we assume that only stars with initial mass $\leq 15$ $M_\odot$ explode successfully, but the results are essentially the same if we use a larger mass. In this case we set $\dot{M}_w(X,t)$ to the value returned by our grid at all times, $M_{\rm SN}(m)$ to the value taken from the tables of \citet{limongi2018} for $m \leq 15$ M$_\odot$, and $M_{\rm SN}(m)$ to zero for $m > 15$ M$_\odot$.


We show our predicted N/O vs.~O/H ratios for the various parameter combinations in \autoref{fig:NO_OH_key_fig}. All the results shown use our fiducial value $v/v_{\rm crit}=0.4$, as discussed in \autoref{model_sec}; we defer discussion of $v/v_{\rm crit} = 0.2$ and $0.6$ to \autoref{rot_dep_sec}. For comparison with the models, we also overplot observed values of N/O versus O/H for Milky Way halo stars \citep{israelian2004, spite2005}, young stars in the Milky Way disk \citep{nieva2012}, and \ion{H}{ii} regions in nearby dwarfs \citep{izotov1999}. 

It is immediately apparent from  \autoref{fig:NO_OH_key_fig} that our first scenario, where low-metallicity dwarf galaxies retain only wind ejecta, produces a distribution of N/O vs.~O/H that is strikingly similar to the observed plateau of constant N/O at low O/H: that is, $\log(\mathrm{N}/\mathrm{O})$ is roughly $-2$ to $-1.5$ independent of the O/H ratio. The reason for this is apparent if one recalls \autoref{fig:IMF_NO}: as soon as stars enter the Wolf-Rayet phase of evolution, the N/O ratio in their winds drops to a low, nearly-constant value. This value is nearly independent of the initial metallicity of the stellar population (which we have seen above is a consequence of the N having both primary and secondary origin).  The second scenario, where galaxies retain SN ejecta, however produces a higher value of N/O ( $\log(\mathrm{N}/\mathrm{O}) \sim -1.2$ to $-1.0$) than the observed distribution. While this likely rules out SN as a the main producers of N in low-metallicity galaxies, it is suggestive that the observed upturn of N/O towards larger values of O/H ($12+\log(\mathrm{O}/\mathrm{H}) \gtrsim 8.0$) may in part be due to larger and more metal-rich galaxies retaining some of their SN ejecta, along with the onset of secondary N from AGB stars at high metallicities.

The dependence of the final abundances on $\epsilon_{\rm p}$ and $\epsilon_*$ can be understood fairly easily. Adding primordial gas changes O/H without altering N/O, causing points to slide horizontally in \autoref{fig:NO_OH_key_fig}. Thus the effect of changing $\epsilon_{\rm p}$, the parameter that controls the amount of primordial gas, is simply to slide points horizontally in \autoref{fig:NO_OH_key_fig}. Thus the horizontal spread in the N/O vs.~O/H plot can be interpreted, not surprisingly, as simply reflecting the amount of primordial versus astrated material in a given star or \ion{H}{ii} region. The effects of $\epsilon_*$ are more subtle: higher $\epsilon_*$ both increases O/H and decreases N/O. The increase in O/H is simply a result of higher star formation efficiency leading to more material being astrated, and thus to a higher total amount of O. The decrease in N/O with $\epsilon_*$ can be understood by reference to \autoref{fig:IMF_NO}. Pure stellar wind ejecta, and SN ejecta for stars with initial Fe abundance $[\mathrm{Fe}/\mathrm{H}]\lesssim -2$, have a very low N/O ratio, close to $\log(\mathrm{N}/\mathrm{O})=-2$. If $\epsilon_* = 1$, then the N and O that we observe are pure stellar ejecta, and thus have the low N/O ratio reflective of that origin. If $\epsilon_* < 1$, on the other hand, then the N and O represent an admixture of stellar wind ejecta and whatever N and O were present in the reservoir prior to star formation, which, per our assumed empirical scaling (which in turn likely reflects the average origin of N and O, including both stellar winds, SNe at a range of metallicities, and intermediate mass stars), has a somewhat larger N/O ratio. Thus N/O increases for $\epsilon_* < 1$.

Taken together, \autoref{fig:NO_OH_key_fig} suggests the following scenario for the origin of the N/O vs.~O/H distribution: at low O/H, we observe a low value of N/O that is independent of O/H on average, but with a large scatter. The low mean value reflects the fact that the N and O we observe at these metallicities originates predominantly in the pre-SN winds of massive stars that were retained in their host galaxy and rapidly mixed back into the ISM, while N released in SN ejecta was mostly lost to the IGM. The large scatter in N/O represents scatter in how much of the N and O are from stellar winds, versus other sources that produce higher or similar N/O ratios depending on metallicity -- the stars or \ion{H}{ii} regions with the lowest N/O are those for which the material is purely massive star wind ejecta, while those with low O/H but higher N/O include a higher proportion of SN ejecta in metal-rich environments ([Fe/H] $\geq -2.0$) or perhaps N produced from intermediate mass stars. At higher O/H, galaxies tend to be larger, and retain a larger proportion of their SN ejecta. Consequently, the mean N/O ratio close to the transition point between low to high O/H favours the value produced by SNe, and at higher O/H ($12+\log(O/H) \gtrsim 8.0$) it may be a combination of SNe and secondary N from AGB stars.
\begin{figure}
\centerline{
\includegraphics[width=0.5\textwidth]{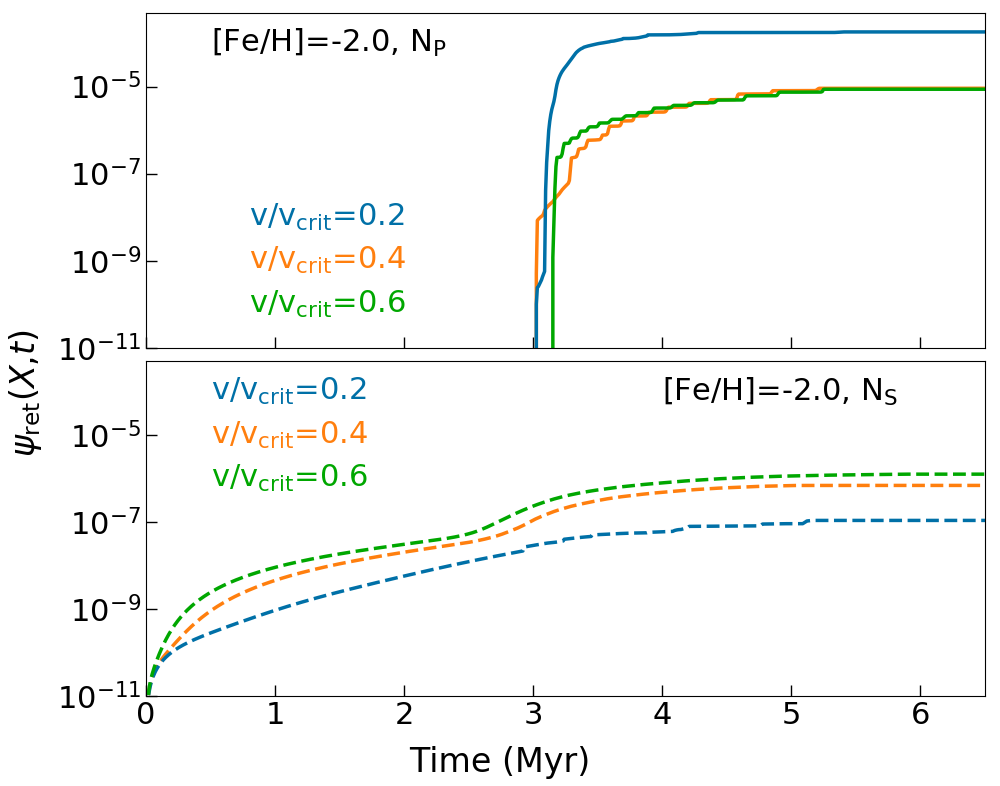}
}

\caption{Time evolution of cumulative mass return $\psi_{\rm{ret}} (X,t)$ purely from winds ($\psi_{\rm{SN}} (X,t)=0.0$) for  N$_P$ (top panel, solid lines) and N$_S$ (bottom panel, dashed lines) for [Fe/H] = -2.0 ([$\alpha$/Fe]=0.4) for three rotation rates, $v/v_{\rm{crit}} = 0.2$ (blue), 0.4 (orange), 0.6 (green). 
}
\label{fig:NP_NS_time_vvcrit_comp}
\end{figure}

\begin{figure}
\centerline{
\includegraphics[width=0.5\textwidth]{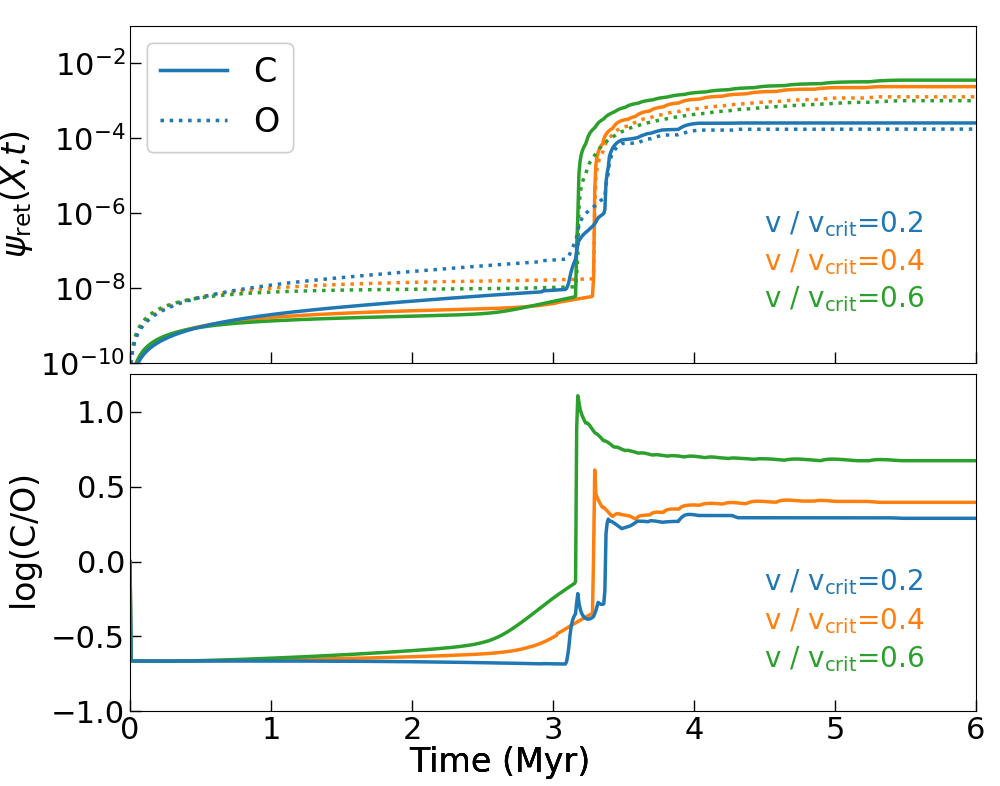}
}

\caption{{\it Top panel:} Time evolution of $\psi_{\rm{ret}} (X,t)$ purely from winds ($\psi_{\rm{SN}} (X,t)=0.0$) of C and O purely from stellar winds for three rotation rates, $v/v_{\rm{crit}} = 0.2$ (blue), $0.4$ (orange) and $0.6$ (green), same as the top panel of \autoref{fig:N_O_time_vvcrit_comp}.  {\it Bottom panel:} $\log$(C/O) versus time, similar to the bottom panel of \autoref{fig:N_O_time_vvcrit_comp}. Both the panels are for [Fe/H]=-2.0 ([$\alpha$/Fe]=0.4).}
\label{fig:C_O_time_vvcrit_comp}
\end{figure}

\begin{figure}
\centerline{
\includegraphics[width=0.5\textwidth]{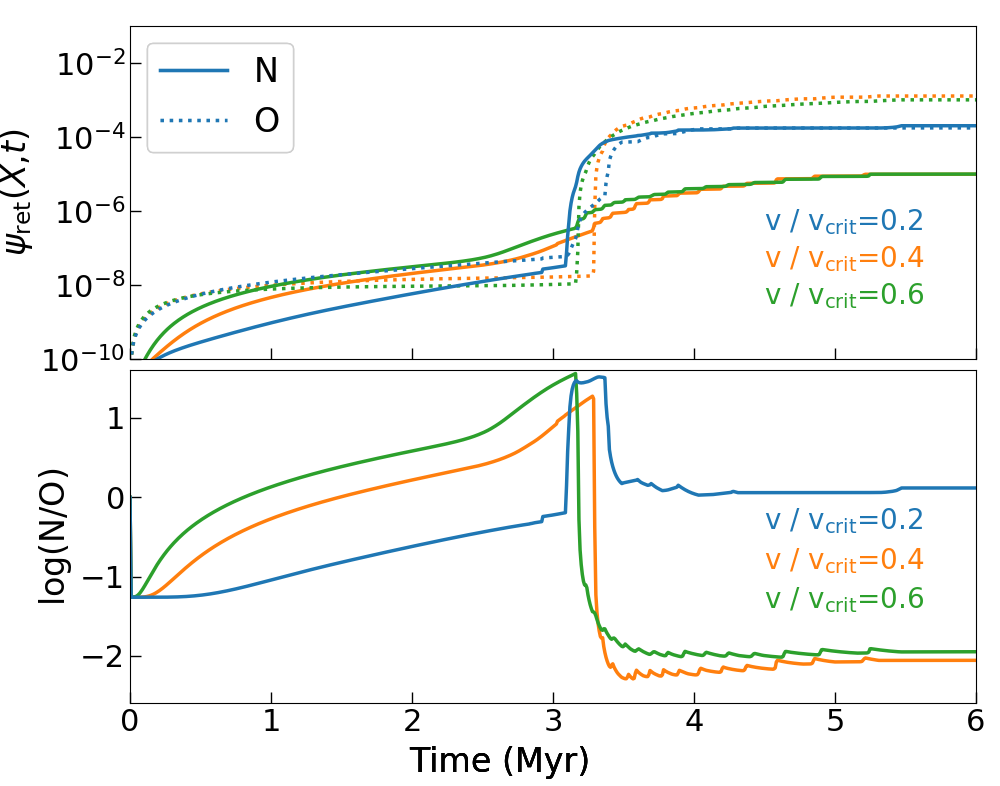}
}

\caption{{\it Top panel:} Time evolution of $\psi_{\rm{ret}} (X,t)$ of N and O purely from stellar winds ($\psi_{\rm{SN}} (X,t)=0.0$) for three rotation rates, $v/v_{\rm{crit}} = 0.2$ (blue), $0.4$ (orange) and $0.6$ (green). The solid and dotted lines represent N and O respectively, similar to top panels of \autoref{fig:IMF_NO}.  {\it Bottom panel:} $\log$(N/O) versus time, similar to the bottom panels of \autoref{fig:IMF_NO}, for the same rotation rates as the top panel. Both the panels are for [Fe/H]=-2.0 ([$\alpha$/Fe]=0.4).
}
\label{fig:N_O_time_vvcrit_comp}
\end{figure}

\begin{figure*}
\centerline{
\includegraphics[width=1.0\textwidth]{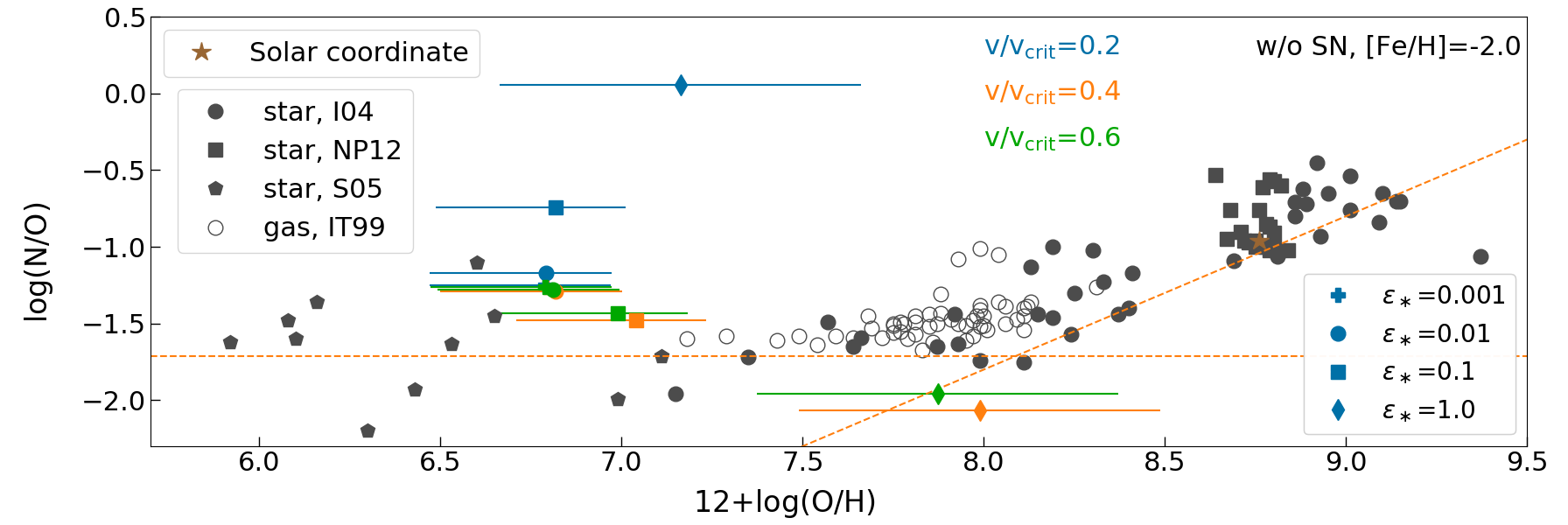}
}

\caption{
Same as the top panel of \autoref{fig:NO_OH_key_fig}, except that the theoretical models shown all have the same metallicity ($[{\rm Fe}/{\rm H}]=-2$, $[\alpha/{\rm Fe}]=0.4$), but have differing rotation rates of $v/v_{\rm crit} = 0.2$ (blue), 0.4 (orange), and 0.6 (green).
}
\label{fig:NO_OH_vvcrit_comp}
\end{figure*}
\subsection{Dependence of N/O on stellar rotation rates}
\label{rot_dep_sec}


Having discussed the origin of N in the context of our fiducial choice of stellar rotation rate, in this section we explore the dependence of our results on this choice.\footnote{In this section, we use a more limited grid that only covers masses from $50-130$ $M_\odot$, because outside this mass range we are unable to complete the MESA calculations at the highest rotation rate, $v/v_{\rm crit} = 0.6$. At this rotation rate, we find that cases with larger initial mass reach the Eddington limit and thus can no longer be treated as being in mechanical equilibrium before reaching the end of core He burning. Conversely, those with starting mass $<50$ $M_\odot$ do not become non-hydrostatic, but require such small time steps that continuing the run all the way to the end of core He burning becomes prohibitively expensive. However, this reduction in mass range has minimal effects on our total yields, since N and O production via winds are completely dominated by stars within the mass range we are able to cover. Thus our reduced grid should not qualitatively affect our conclusions.} To do so, in \autoref{fig:N_O_time_vvcrit_comp} we plot the time evolution of N and O-yields, and the N/O ratio, (similar to the top and the bottom panels of \autoref{fig:IMF_NO} respectively) from stellar winds for three values of $v/v_{\rm {crit}}$ (0.2, 0.4, 0.6) and for $[\mathrm{Fe}/\mathrm{H}]= -2.0$. We separate the N produced in these models into primary and secondary, following the same procedure outlined in \autoref{time_yield_sec} and used in \autoref{fig:NP_NS}, in \autoref{fig:NP_NS_time_vvcrit_comp}.

The main conclusion to be drawn from these two figures is that the $v/v_{\rm crit} = 0.4$ and $0.6$ cases are nearly identical, while the $v/v_{\rm crit} = 0.2$ is qualitatively very different. It is characterised by lower N production at early times, much higher N and lower O production at later times, and a much larger ratio of primary to secondary N. We can understand these changes as resulting from shifts in the nuclear burning regime that supplies the elements that are ultimately ejected in the stellar wind. As shown in \citet{roy2019}, return of secondary N to the ISM by a rapidly-rotating massive star starts while the star is still on the main sequence, due to both rotational dredge-up and exposure of core material by weak main sequence winds. Thus the main nucleosynthetic process feeding the wind at early times is the CNO cycle. However, once core H is exhausted and core He burning starts, the O abundance in stellar cores rapidly rises, and N production switches to primary. At this point the main region feeding the wind is undergoing He rather than CNO burning, and this explains the sharp drop in N/O ratio and sharp increase in primary N that is visible at $\approx 3.5$ Myr in the figures. By contrast, in a more slowly-rotating star, rotational dredge up and mass loss during the main sequence are both much weaker, so less secondary N is returned to the ISM. Moreover, the weaker mass loss and rotational mixing also mean that, once core H is exhausted, there remains a substantial shell of H-rich material outside the core that has not undergone any nuclear processing. While the core is burning He, this shell undergoes CNO burning, and the CNO-burning shell rather than the He-burning core provides most of the material that is ultimately lost in winds. The shell contains much less O and much more primary N than the He-burning core that dominates production in the more rapidly rotating case. We can confirm that this explanation is correct by examining \autoref{fig:C_O_time_vvcrit_comp} which shows the time evolution of C and O, and their ratio in \autoref{fig:C_O_time_vvcrit_comp}, in the same manner as \autoref{fig:N_O_time_vvcrit_comp}. We see that ratio of C to O, both of which are primary elements is much less sensitive to rotation rate than the N/O ratio, and that the sense of dependence is reversed compared to N/O: as the rotation rate increases, the C/O ratio also increases, the opposite of the N/O trend. Thus as we go from slowly- to rapidly-rotating stars, we see that the ejecta during the WR phase contain more and more C, and less and less N, as expected if the ejecta are undergoing more and more CNO-cycling.

We explore the consequences of changing the rotation rate for chemical abundances in \autoref{fig:NO_OH_vvcrit_comp}, which is constructed in the same manner as the top panel of \autoref{fig:NO_OH_key_fig}, but for runs with varying rotation rates; to avoid clutter in this figure, we show only the case $[{\rm Fe}/{\rm H}]=-2$. We see from the plot that, while our cases with $v/v_{\rm crit} = 0.4$ and 0.6 provide good matches to the observations, calculations with $v/v_{\rm crit} = 0.2$ do not, because they produce N/O ratios that are much too high. We therefore conclude that massive stellar winds are a viable explanation for the observed distribution of N and O in dwarf galaxies only if these stars are typically moderate- to fast-rotators at birth.



\section{Discussions \& Conclusions}
\label{disc_sec}

Thus far we have shown that nucleosynthesis in the winds of rotating massive stars produces amounts of nitrogen, and ratios of nitrogen to oxygen, that are consistent with the abundance patterns observed in metal-poor stars and dwarf galaxies. In this section, we summarise our primary findings, and use them to propose a scenario for N production in the early Universe.

We begin by summarising the three most salient findings from our model grid.
\begin{itemize}
\item The winds of metal-poor ($[\mathrm{Fe}/\mathrm{H}\lesssim -2]$), massive ($\gtrsim 10$ $M_\odot$) stars born with relatively rapid rotation ($v/v_{\rm crit}\gtrsim 0.4$) contain substantial amounts of secondary nitrogen, which is ejected while the stars are on the main sequence, and primary nitrogen, which is produced during and immediately after the onset of core He burning. The winds of these stars also contain oxygen, and the N/O ratio of the wind is in the range $\log\left(\mathrm{N}/\mathrm{O}\right) \approx -2$ to $-1.5$, independent of total oxygen metallicity or production. This is very similar to the N/O ratio measured in Milky Way halo stars and in dwarf galaxies with $12+\log(\mathrm{O}/\mathrm{H})\leq 8$.

\item While the mean N/O ratio produced by winds is a good match to the mean observed value, there is also significant scatter in both the data and the models. In the models, this scatter is a result of variation in both the metallicity of the stellar population and in the amount of non-astrated gas with which those stellar winds mix. Variations in these quantities plausibly explains the observed high level of variation in N/O ratio at low O/H.

\item SNe produce a higher N/O ratios ($\log\left(\mathrm{N}/\mathrm{O}\right) \approx -1$) compared to winds, and SN ejecta are much faster than stellar wind ejecta: $\geqslant 3,500$ km s$^{-1}$ versus $\sim 1,000$ km s$^{-1}$ (O stars) and $\sim 200$ km s$^{-1}$ (WR stars). This difference translates to a three-order of magnitude difference in post-shock temperature of SN ejecta compared to WR wind-ejecta, so that shocked stellar wind material is much more likely to be able to cool and be retained by a galaxy than SN ejecta.
\end{itemize}

Taken together, these findings allow us to sketch out a scenario for the origin of nitrogen in the low-metallicity Universe, and an explanation for the observed scaling of N/O with O/H. Star formation at low metallicity primarily takes place in metal-poor dwarf galaxies. These have shallow potential wells, and simulations of such galaxies
 \citep{emerick2018,  emerick2019} 
indicate that SNe outflows easily escape them. Indeed, preferential metal ejection is required in order to explain the mass-metallicity relation in dwarf galaxies \citep{peeples2011, peeples2013, forbes2019}. 
Consequently, the initial buildup of metals in these galaxies is driven not primarily by SNe, but by stellar wind ejecta. The winds are much easier to retain due to their lower velocities, and shocked wind material can mix with ambient ISM and form stars even before the bulk of SNe explode \citep{lochhaas2017}. The observed abundances of N and O in systems with $12+\log(\mathrm{O}/\mathrm{H}) \lesssim 8$ mostly reflect this origin channel, whose characteristics are a mean $\log(\mathrm{N}/\mathrm{O})$ ratio of $\approx -1.5$ to $-2$, and a wide scatter, caused by variation in the initial metallicity of the stellar population, and in the amount of mixing between stellar winds and ambient gas. Over time, the mostly-primary N produced by massive star winds is supplemented by secondary N produced in AGB stars, with longer evolution times. As a result the N/O ratio begins to rise to the values of $\approx -0.5$ to $-1$ that characterise Solar metallicity galaxies.

\section*{Acknowledgments}
We thank the referee Paola Marigo for insightful comments to improve our paper. AR acknowledges the IMPULSION grant (grant code: IDEX/IMP/2020/08) from IDEX Universit\'e de Lyon. AR acknowledges MD for initiating this project and tributes this work to his memory. AR and MRK acknowledge support from the Australian Research Council's \textit{Discovery Projects} and \textit{Future Fellowship} funding scheme, awards DP190101258 and FT180100375. LK and AR acknowledge the support of Australian Research Council's \textit{Laureate Fellowship} award FL150100113. MD and RS acknowledge the support of the Australian Research Council (ARC) through Discovery project DP16010363. Parts of this research were conducted by the Australian Research Council Centre of Excellence for All Sky Astrophysics in 3 Dimensions (ASTRO 3D), through project number CE170100013. AH has been supported, in part, by a grant from Science and Technology Commission of Shanghai Municipality (Grants No.16DZ2260200) and National Natural Science Foundation of China (Grants No.11655002), and by the Australian Research Council Centre of Excellence for Gravitational Wave Discovery (OzGrav), through project number CE170100004. This work benefited from support by the National Science Foundation under Grant No. PHY-1430152 (JINA Center for the Evolution of the Elements). This research/project was undertaken with the assistance of resources and services from the National Computational Infrastructure (NCI), which is supported by the Australian Government.

\section*{Data availability}
The data underlying this article will be shared on request to the corresponding author (AR).

\bibliographystyle{mnras}
\bibliography{refs}




\appendix

\section{Dependence of yields and N/O on the upper mass-limit of SNe}
\label{upper_mass_sec}
\begin{figure*}
\centerline{
\includegraphics[width=1.0\textwidth]{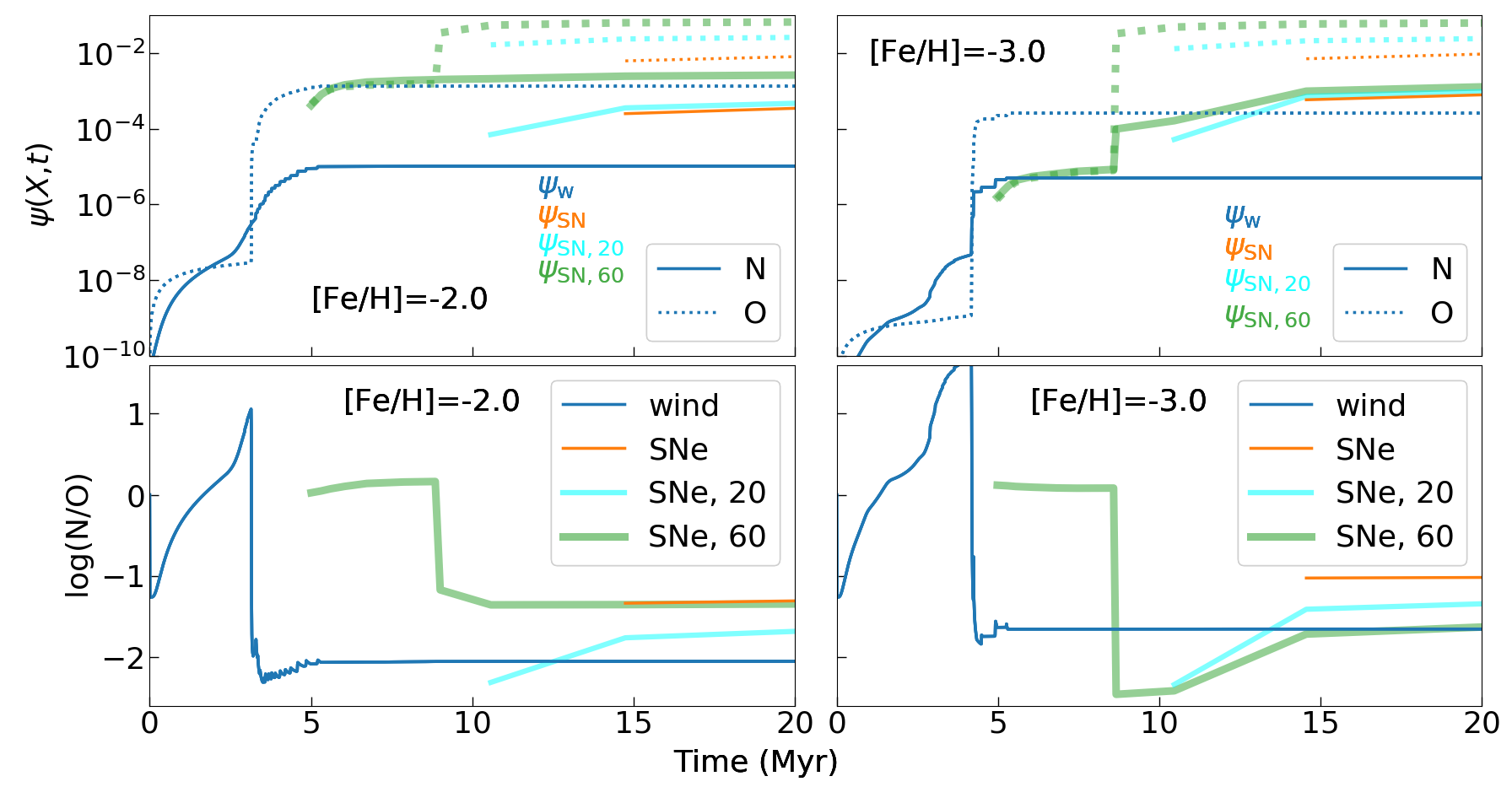}
}

\caption{Same as \autoref{fig:IMF_NO}, except here we show the SNe cut-off limit for two more masses, 20 ($\psi_{{\rm SN},\, 20}$) and 60 M$_\odot$ ($\psi_{{\rm SN},\, 60}$), and their corresponding N/O. 
}
\label{fig:IMF_NO_20_60Msun}
\end{figure*}

\begin{figure*}
\centerline{
\includegraphics[width=1.0\textwidth]{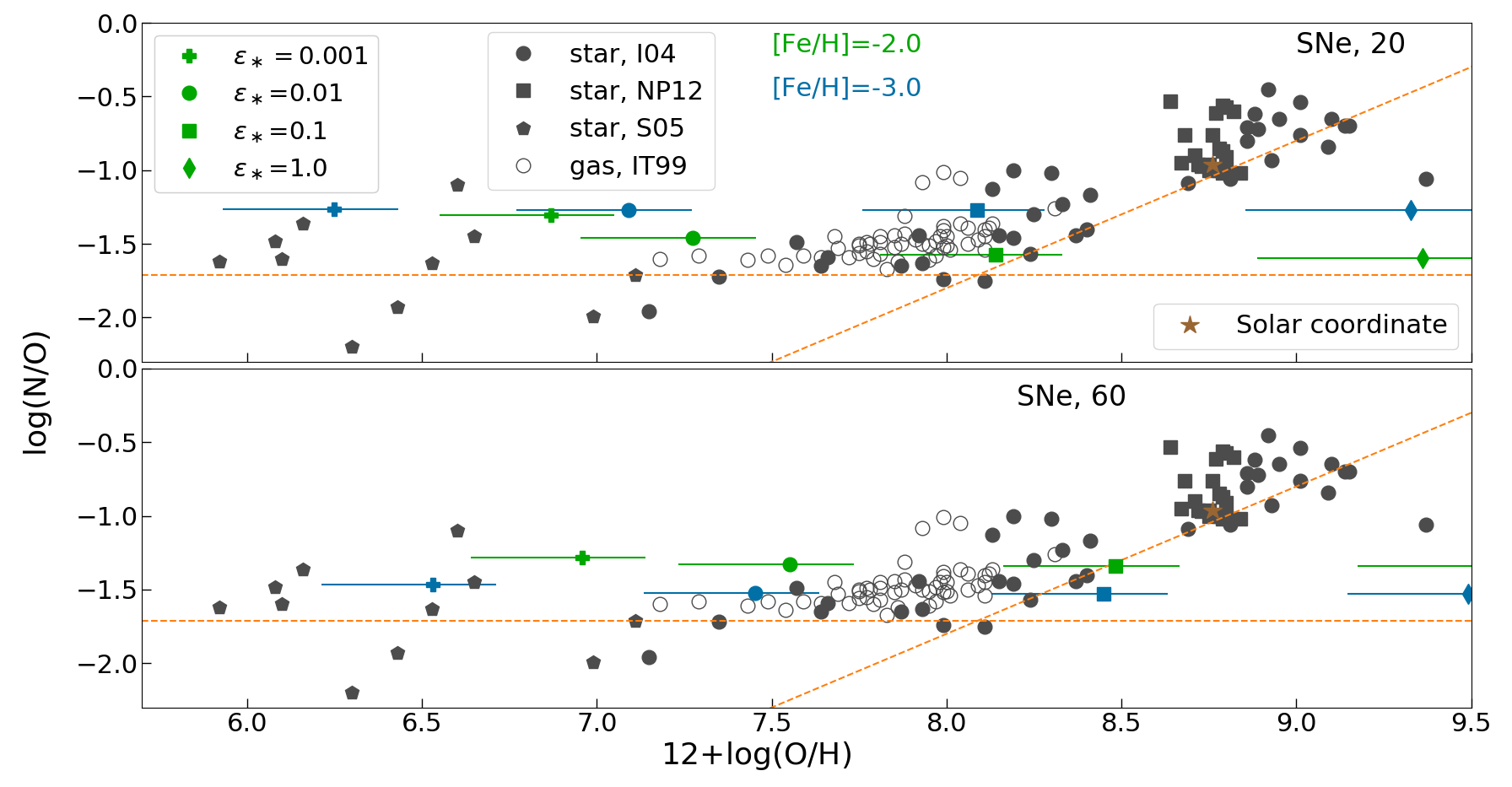}
}

\caption{Same as the bottom panel of \autoref{fig:NO_OH_key_fig}, except that the SNe mass limits are 20 ({\it top panel}) and 60 M$_\odot$ ({\it bottom panel})
}
\label{fig:NO_OH_20_60Msun}
\end{figure*}

In this Appendix, we explore the sensitivity of our results to the upper mass-limit of SN beyond which we assume massive stars to collapse as black holes. In \autoref{fig:IMF_NO_20_60Msun} we show $\psi_{\rm ret} (X,t)$ for O and N, exactly as in \autoref{fig:IMF_NO}, for but assuming that successful SNe occur up to masses of 20 and 60 M$_\odot$, as compared to our fiducial choice 15 M$_\odot$. We first discuss the 20 $M_\odot$ case, which the figure shows does not change the N yields compared to 15 M$_\odot$ limit, although O yields increase by a factor of 4 for both metallicities ([Fe/H]$=-2.0\, , -3.0$). Therefore, $\log(N/O)$ for 20 M$_\odot$ decreases to $\sim -1.4$ from $-1.0$ (for 15 M$_\odot$). However, $\log(N/O) \sim -1.4$ is still substantially above the observed plateau value, which is $\sim -2.0$, and above the value produced by winds. We illustrate how this affects the location of stars in the O/H - N/O plane in the top panel of \autoref{fig:NO_OH_20_60Msun}. As is clear from the figure, the points for SN production with a 20 M$_\odot$ limit are still at systematically higher N/O than the majority of the observations, and, moreover, the $12+\log((\mathrm{O/H})$ is also substantially higher, lying closer to the crossover at $12+\log((\mathrm{O/H}) \sim 7.5$ than to the plateau region at lower $12 + \log(\mathrm{O/H})$ unless we assume exceptionally small star formation efficiency, $\epsilon_* \sim 10^{-3}$.

For an even higher mass-limit of 60 M$_\odot$, the change in both N and O yields is drastic for higher metallicity, [Fe/H]$=-2.0$. This results in $\log(\textbf{N/O}) \sim -1.0$, similar to the result for a 15 M$_\odot$ limit (see both top and bottom panels of the leftmost panels of \autoref{fig:IMF_NO_20_60Msun}). However, for lower metallicity, [Fe/H]$=-3.0$ (the rightmost panels of \autoref{fig:IMF_NO_20_60Msun}), the N yield does not differ from 15 and 20 M$_\odot$ limits, although the O yield differs (increases) significantly resulting in $\log(\mathrm{N/O}) \sim -2.0$, similar to the value for winds. The same results are reflected in the bottom panel of \autoref{fig:NO_OH_20_60Msun}. It is conceivable that SNe could explain the observed N/O ratio \textit{if} Type II SNe can occur for stars as massive as 60 M$_\odot$, and if the ejecta can be retained in the galaxy. However, we caution that both of these are extreme assumptions, as modern models suggest that Type II SNe are unlikely for stars with initial masses as large as 60 M$_\odot$ \citep{sukhbold16}.  


\bsp	
\label{lastpage}
\end{document}